\def\eck#1{\left\lbrack #1 \right\rbrack}
\def\eckk#1{\bigl[ #1 \bigr]}
\def\rund#1{\left( #1 \right)}
\def\abs#1{\left\vert #1 \right\vert}

\def\ave#1{\left\langle #1 \right\rangle}

\def\part#1#2{{\partial #1\over\partial #2}}

\def\Re{{\cal R}\hbox{e}}

\def\A{{\cal A}}

\def\O{{\cal O}}

\def\d{{\rm d}}

\def\eps{{\epsilon}}

\def\vp{\varphi}
\def\vt{{\vartheta}}

\def\Real{{\rm I\mathchoice{\kern-0.70mm}{\kern-0.70mm}{\kern-0.65mm}%
  {\kern-0.50mm}R}}
\def\C{\rm C\kern-.42em\vrule width.03em height.58em depth-.02em
       \kern.4em}
\font \bolditalics = cmmib10
\def\bx#1{\leavevmode\thinspace\hbox{\vrule\vtop{\vbox{\hrule\kern1pt
        \hbox{\vphantom{\tt/}\thinspace{\bf#1}\thinspace}}
      \kern1pt\hrule}\vrule}\thinspace}

\def \vc #1{{\textfont1=\bolditalics \hbox{$\bf#1$}}}
{\catcode`\@=11
\gdef\SchlangeUnter#1#2{\lower2pt\vbox{\baselineskip 0pt \lineskip0pt
  \ialign{$\m@th#1\hfil##\hfil$\crcr#2\crcr\sim\crcr}}}
}

\def\ueber#1#2{{\setbox0=\hbox{$#1$}%
  \setbox1=\hbox to\wd0{\hss$\scriptscriptstyle #2$\hss}%
  \offinterlineskip
  \vbox{\box1\kern0.4mm\box0}}{}}

\def\bx#1{\leavevmode\thinspace\hbox{\vrule\vtop{\vbox{\hrule\kern1pt
        \hbox{\vphantom{\tt/}\thinspace{\bf#1}\thinspace}}
      \kern1pt\hrule}\vrule}\thinspace}

\voffset=0pt

\magnification=\magstep1
\input epsf
\voffset= 0.0 true cm
\vsize=19.8 cm     
\hsize=13.5 cm
\hfuzz=2pt
\tolerance=500
\abovedisplayskip=3 mm plus6pt minus 4pt
\belowdisplayskip=3 mm plus6pt minus 4pt
\abovedisplayshortskip=0mm plus6pt
\belowdisplayshortskip=2 mm plus4pt minus 4pt
\predisplaypenalty=0
\footline={\tenrm\ifodd\pageno\hfil\folio\else\folio\hfil\fi}

\def\la{\mathrel{\hbox{\rlap{\hbox{\lower4pt\hbox{$\sim$}}}\hbox{$<$}}}}
\def\ga{\mathrel{\hbox{\rlap{\hbox{\lower4pt\hbox{$\sim$}}}\hbox{$>$}}}}

\def\utw{\smash{\rlap{\lower5pt\hbox{$\sim$}}}}
\def\udtw{\smash{\rlap{\lower6pt\hbox{$\approx$}}}}

\def\getsto{\mathrel{\hbox{\rlap{$\gets$}\hbox{\raise2pt\hbox{$\to$}}}}}
\def\lid{\mathrel{\hbox{\rlap{\hbox{\lower4pt\hbox{$=$}}}\hbox{$<$}}}}
\def\gid{\mathrel{\hbox{\rlap{\hbox{\lower4pt\hbox{$=$}}}\hbox{$>$}}}}
\def\sol{\mathrel{\hbox{\rlap{\hbox{\raise4pt\hbox{$\sim$}}}\hbox{$<$}}}
}
\def\sog{\mathrel{\hbox{\rlap{\hbox{\raise4pt\hbox{$\sim$}}}\hbox{$>$}}}
}
\def\lse{\mathrel{\hbox{\rlap{\hbox{\raise4pt\hbox{$<$}}}\hbox{$\simeq$}
}}}
\def\gse{\mathrel{\hbox{\rlap{\hbox{\raise4pt\hbox{$>$}}}\hbox{$\simeq$}
}}}
\def\grole{\mathrel{\hbox{\lower2pt\hbox{$<$}}\kern-8pt
\hbox{\raise2pt\hbox{$>$}}}}
\def\leogr{\mathrel{\hbox{\lower2pt\hbox{$>$}}\kern-8pt
\hbox{\raise2pt\hbox{$<$}}}}
\def\loa{\mathrel{\hbox{\rlap{\hbox{\lower4pt\hbox{$\approx$}}}\hbox{$<$
}}}}
\def\goa{\mathrel{\hbox{\rlap{\hbox{\lower4pt\hbox{$\approx$}}}\hbox{$>$
}}}}

%
%

\font\kleinhalbcurs=cmmib10 scaled 833
\font\eightrm=cmr8
\font\sixrm=cmr6
\font\eighti=cmmi8
\font\sixi=cmmi6
\skewchar\eighti='177 \skewchar\sixi='177
\font\eightsy=cmsy8
\font\sixsy=cmsy6
\skewchar\eightsy='60 \skewchar\sixsy='60
\font\eightbf=cmbx8
\font\sixbf=cmbx6
\font\eighttt=cmtt8
\hyphenchar\eighttt=-1
\font\eightsl=cmsl8
\font\eightit=cmti8

\font\bxf=cmbx10
  \mathchardef\Gamma="0100
  \mathchardef\Delta="0101
  \mathchardef\Theta="0102
  \mathchardef\Lambda="0103
  \mathchardef\Xi="0104
  \mathchardef\Pi="0105
  \mathchardef\Sigma="0106
  \mathchardef\Upsilon="0107
  \mathchardef\Phi="0108
  \mathchardef\Psi="0109
  \mathchardef\Omega="010A
\def\rahmen#1{\vskip#1truecm}
\def\begfig#1cm#2\endfig{\par
\setbox1=\vbox{\rahmen{#1}#2}%
\dimen0=\ht1\advance\dimen0by\dp1\advance\dimen0by5\baselineskip
\advance\dimen0by0.4true cm
\ifdim\dimen0>\vsize\pageinsert\box1\vfill\endinsert
\else
\dimen0=\pagetotal\ifdim\dimen0<\pagegoal
\advance\dimen0by\ht1\advance\dimen0by\dp1\advance\dimen0by1.4true cm
\ifdim\dimen0>\vsize
\topinsert\box1\endinsert
\else\vskip1true cm\box1\vskip4true mm\fi
\else\vskip1true cm\box1\vskip4true mm\fi\fi}
\def\figure#1#2{\smallskip\setbox0=\vbox{\noindent\petit{\bf Fig.\ts#1.\
}\ignorespaces #2\smallskip
\count255=0\global\advance\count255by\prevgraf}%
\ifnum\count255>1\box0\else
\centerline{\petit{\bf Fig.\ts#1.\ }\ignorespaces#2}\smallskip\fi}

\def\xfigure#1#2#3#4{\midinsert\noindent
    $$\epsfxsize=#4truecm\epsffile{#3}$$
    \figure{#1}{#2}\endinsert}


\def\begtab#1cm#2\endtab{\par
\ifvoid\topins\midinsert\vbox{#2\rahmen{#1}}\endinsert
\else\topinsert\vbox{#2\kern#1true cm}\endinsert\fi}
\def\rahmen#1{\vskip#1truecm}
\def\begpet{\vskip6pt\bgroup\petit}
\def\endpet{\vskip6pt\egroup}
\def\begref{\par\bgroup\petit
\let\it=\rm\let\bf=\rm\let\sl=\rm\let\INS=N}
\def\petit{\def\rm{\fam0\eightrm}%
\textfont0=\eightrm \scriptfont0=\sixrm \scriptscriptfont0=\fiverm
 \textfont1=\eighti \scriptfont1=\sixi \scriptscriptfont1=\fivei
 \textfont2=\eightsy \scriptfont2=\sixsy \scriptscriptfont2=\fivesy
 \def\it{\fam\itfam\eightit}%
 \textfont\itfam=\eightit
 \def\sl{\fam\slfam\eightsl}%
 \textfont\slfam=\eightsl
 \def\bf{\fam\bffam\eightbf}%
 \textfont\bffam=\eightbf \scriptfont\bffam=\sixbf
 \scriptscriptfont\bffam=\fivebf
 \def\tt{\fam\ttfam\eighttt}%
 \textfont\ttfam=\eighttt
 \normalbaselineskip=9pt
 \setbox\strutbox=\hbox{\vrule height7pt depth2pt width0pt}%
 \normalbaselines\rm
\def\vec##1{\setbox0=\hbox{$##1$}\hbox{\hbox
to0pt{\copy0\hss}\kern0.45pt\box0}}}%
\let\ts=\thinspace
%
\font \tafontt=     cmbx10 scaled\magstep2
\font \tafonts=     cmbx7  scaled\magstep2
\font \tafontss=     cmbx5  scaled\magstep2
\font \tamt= cmmib10 scaled\magstep2
\font \tams= cmmib10 scaled\magstep1
\font \tamss= cmmib10
\font \tast= cmsy10 scaled\magstep2
\font \tass= cmsy7  scaled\magstep2
\font \tasss= cmsy5  scaled\magstep2
\font \tasyt= cmex10 scaled\magstep2
\font \tasys= cmex10 scaled\magstep1
\font \tbfontt=     cmbx10 scaled\magstep1
\font \tbfonts=     cmbx7  scaled\magstep1
\font \tbfontss=     cmbx5  scaled\magstep1
\font \tbst= cmsy10 scaled\magstep1
\font \tbss= cmsy7  scaled\magstep1
\font \tbsss= cmsy5  scaled\magstep1

\newbox\chsta\newbox\chstb\newbox\chstc
\def\centerpar#1{{\advance\hsize by-2\parindent
\rightskip=0pt plus 4em
\leftskip=0pt plus 4em
\parindent=0pt\setbox\chsta=\vbox{#1}%
\global\setbox\chstb=\vbox{\unvbox\chsta
\setbox\chstc=\lastbox
\line{\hfill\unhbox\chstc\unskip\unskip\unpenalty\hfill}}}%
\leftline{\kern\parindent\box\chstb}}
 \def \chap#1{
    \vskip24pt plus 6pt minus 4pt
    \bgroup
 \textfont0=\tafontt \scriptfont0=\tafonts \scriptscriptfont0=\tafontss
 \textfont1=\tamt \scriptfont1=\tams \scriptscriptfont1=\tamss
 \textfont2=\tast \scriptfont2=\tass \scriptscriptfont2=\tasss
 \textfont3=\tasyt \scriptfont3=\tasys \scriptscriptfont3=\tenex
     \baselineskip=18pt
     \lineskip=18pt
     \raggedright
     \pretolerance=10000
     \noindent
     \tafontt
     \ignorespaces#1\vskip7true mm plus6pt minus 4pt
     \egroup\noindent\ignorespaces}%
 \def \sec#1{
     \vskip25true pt plus4pt minus4pt
     \bgroup
 \textfont0=\tbfontt \scriptfont0=\tbfonts \scriptscriptfont0=\tbfontss
 \textfont1=\tams \scriptfont1=\tamss \scriptscriptfont1=\kleinhalbcurs
 \textfont2=\tbst \scriptfont2=\tbss \scriptscriptfont2=\tbsss
 \textfont3=\tasys \scriptfont3=\tenex \scriptscriptfont3=\tenex
     \baselineskip=16pt
     \lineskip=16pt
     \raggedright
     \pretolerance=10000
     \noindent
     \tbfontt
     \ignorespaces #1
     \vskip12true pt plus4pt minus4pt\egroup\noindent\ignorespaces}%
 \def \subs#1{
     \vskip15true pt plus 4pt minus4pt
     \bgroup
     \bxf
     \noindent
     \raggedright
     \pretolerance=10000
     \ignorespaces #1
     \vskip6true pt plus4pt minus4pt\egroup
     \noindent\ignorespaces}%
 \def \subsubs#1{
     \vskip15true pt plus 4pt minus 4pt
     \bgroup
     \bf
     \noindent
     \ignorespaces #1\unskip.\ \egroup
     \ignorespaces}
\def\footnoterule{\kern-3pt\hrule width 2true cm\kern2.6pt}
\newcount\footcount \footcount=0
\def\advftncnt{\advance\footcount by1\global\footcount=\footcount}
\def\fonote#1{\advftncnt$^{\the\footcount}$\begingroup\petit
       \def\textindent##1{\hang\noindent\hbox
       to\parindent{##1\hss}\ignorespaces}%
\vfootnote{$^{\the\footcount}$}{#1}\endgroup}

\newcount\sterne
\outer\def\byebye{\bigskip\typeset
\sterne=1\ifx\speciali\undefined\else
\bigskip Special caracters created by the author
\loop\smallskip\noindent special character No\number\sterne:
\csname special\romannumeral\sterne\endcsname
\advance\sterne by 1\global\sterne=\sterne
\ifnum\sterne<11\repeat\fi
\vfill\supereject\end}
\def\typeset{\centerline{\petit This article was processed by the author
using the \TeX\ Macropackage from Springer-Verlag.}}
\def\s{{({\rm s})}}
\def\dder#1#2{{\d^2 #1\over \d #2^2}}
\def\M{{\cal M}}
\def\m{M_{\rm ap}}
\chap{\centerline{A new measure for cosmic shear}}
\bigskip
\centerline{\bf Peter Schneider, Ludovic van
Waerbeke, Bhuvnesh Jain \& Guido Kruse}
\medskip
\centerline{\bf Max-Planck-Institut f\"ur Astrophysik}
\centerline{\bf Postfach 1523}
\centerline{\bf D-85740 Garching, Germany}
\bigskip
\sec{Abstract}
Cosmic shear, i.e., the distortion of images of high-redshift galaxies
through the tidal gravitational field of the large-scale matter
distribution in the Universe, offers the opportunity to measure the
power spectrum of the cosmic density fluctuations without any
reference to the relation of dark matter to luminous tracers. We
consider here a new statistical measure for cosmic shear, the aperture
mass $\m(\theta)$, which is defined as a spatially filtered projected
density field and which can be measured
directly from the image distortions of high-redshift galaxies. By
selecting an appropriate spatial filter function, the dispersion of
the aperture mass
is a convolution of the power spectrum of the
projected density field with a narrow kernel, so that $\ave{\m(\theta)}$
provides a well localized estimate of the power spectrum at
wavenumbers $s\sim 5/\theta$. We calculate $\ave{\m^2}$ for various
cosmological models, using the fully non-linear power spectrum of the
cosmic density fluctuations. The non-linear evolution yields a
significant increase of $\ave{\m^2}$ relative to the linear growth
on scales below $\sim 1/2$\ts degree.  

The third-order moment of $\m$ can be used to define a skewness,
which is a measure of the non-Gaussianity of the density field. We
present the first calculation of the skewness of the shear 
in the frame of quasi-linear theory of
structure growth. We show that it yields a sensitive measure of the
cosmological model; in particular, it is independent of the
normalization of the power spectrum. 

Several practical estimates for $\ave{\m^2}$ are constructed and their
dispersions calculated. On scales below a few arcminutes, the
intrinsic ellipticity distribution of galaxies is the dominant source
of noise, whereas on larger scales, the cosmic variance becomes the
most important contribution. We show that measurements of $\m$ in two
adjacent apertures are virtually uncorrelated, which implies that 
an image with side-length $L$ can yield $\eckk{L/(2\theta)}^2$
mutually independent estimates for $\m$. We show that one square
degree of a high-quality image is sufficient to detect the cosmic
shear with the $\m$-statistics on scales below $\sim 10$\ts arcmin, and to
estimate its amplitude with an accuracy of $\sim 30$\% on scales below
$\sim 5$\ts arcmin.

\vfill\eject

\sec{1 Introduction}
Gravitational light deflection caused by an inhomogeneous distribution
of matter in the Universe causes observable effects on the images of
distant sources. Whereas mass concentrations on scales of galaxies and
clusters yield strong lensing effects -- multiple images, (radio)
rings, and giant luminous arcs -- density inhomogeneities on larger
scales or with less concentration causes weak lensing effects. In
particular, the shape and observable flux of distant galaxies is
affected by the tidal component of the gravitational field and the
density fluctuations along their lines-of-sight, respectively. Whereas
these lensing effects are too weak to be detected in individual galaxy
images, they can be investigated statistically. Assuming that the
intrinsic orientations of galaxies are random, cosmic shear -- i.e., the
line-of-sight integrated tidal gravitational field -- can be detected
as a net alignment of galaxy images on a given patch of the sky.

The statistical properties of the cosmic shear are directly linked to
the statistical properties of the density inhomogeneities (Gunn 1967,
Blandford \& Jaroszy\'nski 1981). In particular, any two-point
statistics, like the two-point correlation function of galaxy image
ellipticities or the mean quadratic image ellipticity, can be
expressed as a redshift integral over the power spectrum of the
cosmological density fluctuations, weighted by geometrical factors
depending on the source redshift distribution
(Blandford et al.\ts 1991; Miralda-Escud\'e 1991; Kaiser 1992, 1996,
hereafter K96;
Villumsen 1996). These geometrical factors, as well as the redshift
evolution of the power spectrum, depend on the cosmological
model. Therefore, a quantitative analysis of the cosmic shear
statistics can provide strong constraints on cosmological parameters
and the shape of the fluctuation power spectrum. 

It should be stressed that this approach to determine the density
fluctuations in the Universe does not rely on assumptions about the
relation between luminous and dark matter. It is therefore of
comparable interest as the investigation of the cosmic microwave
background radiation. In fact, these two approaches are complementary:
Whereas the CMB measures the fluctuation spectrum at the time of
recombination, where the density perturbations were small, the cosmic
shear probes the fluctuation spectrum at relatively small redshifts,
$z\sim 0.5$, where it has become non-linear on scales smaller than
$\sim 5$\ts Mpc. Secondly, due to the finite thickness of the
recombination shell, the CMB is expected to show no structure on
angular scales below $\sim 5$\ts arcmin, corresponding to comoving
scales of $\sim 10$\ts Mpc, whereas cosmic shear can probe the density
fluctuations also on much smaller scales. Finally, a comparison of the
results from both techniques can provide a beautiful confirmation of
the gravitational instability theory of structure growth.

The above-mentioned authors have calculated the two-point correlation
function of galaxy image ellipticities and the rms image ellipticity
using the linear theory of density instabilities. On angular scales
below $\sim 20$\ts arcmin, one starts to probe the density fluctuation
spectrum on scales below $\sim 5 h^{-1}$\ts Mpc (where $h$ is the
current value of the Hubble constant in units of 100\ts km/s/Mpc)
where the density fluctuations are nonlinear. Jain \& Seljak (1997;
hereafter JS) generalized the previous investigations of cosmic shear
to the fully non-linear evolution of the power spectrum, and found
that the expected rms shear increases by about a factor of two on
small scales, relative to the predictions from the linear evolution of
the power spectrum. Given that the expected rms shear on arcminutes
scale is of the order of a few percent, this enhancement dramatically
increases the possibility to detect cosmic shear with currently
existing instruments and data analysis techniques. In fact, deep
observations of fields around radio QSOs have revealed the presence of
a coherent shear pattern (Fort et al.\ts 1996, Bower \& Smail 1997)
which has been interpreted as cosmic shear (Schneider et al.\ts 1997),
though the selection of these fields was based on the magnification
bias hypothesis (Bartelmann \& Schneider 1994, Ben\'\i tez \& Mart\'\i
nez-Gonz\'alez 1997, and references therein) for these luminous radio
QSOs. Therefore, these measurements should not be interpreted in any
statistical sense; nonetheless, they have shown that a shear of a few
percent is measurable even on scales as small as 2\ts arcmins.

An attempt to measure cosmic shear on a single $\sim 9$\ts arcmin
field did not yield a significant signal (Mould et al.\ts 1994),
though a reanalysis of the same data with a somewhat less conservative
approach found a fairly high significance (Villumsen 1995). The
development of wide-field cameras and sophisticated data analysis
methods specifically targeted at weak lensing studies (Bonnet \&
Mellier 1995; Kaiser, Squires \& Broadhurst 1995; Luppino \& Kaiser
1996; van Waerbeke et al.\ts 1997) suggest that the discovery of
cosmic shear in random fields, and 
its quantitative analysis, are just lurking around the corner.

Higher-order than two-point statistics for cosmic shear are difficult to
estimate analytically. Bernardeau, van Waerbeke \& Mellier (1997;
hereafter BvWM) have calculated the skewness of the projected surface
mass density field, using the quasi-linear theory of structure growth.
They pointed out that the skewness is a powerful indicator for the
cosmic density parameter; in particular, it is independent of the
normalization of the power spectrum, and fairly independent
of the cosmological constant. A practical difficulty related
to that measure is that the projected surface mass density is less
directly linked with observables. Whereas it is tightly related to the
magnification of sources, this by itself is difficult to observe.
One would therefore like to measure the skewness of a
quantity that is directly related to the shear, which is observable from
image ellipticities.

Such a quantity is provided by the $\m$-statistics, introduced by
Kaiser et al.\ts (1994) and Schneider (1996), as a generalization of
the $\zeta$-statistics introduced by Kaiser (1995). The latter yields
an unbiased estimate of the mean surface mass density within a circle,
minus the mean surface mass density within an annulus surrounding this
circle, and can be obtained from the shear within the annulus.  First
applied to the cluster MS1224 (Fahlman et al.\ 1994), this ``aperture
densitometry'' has yielded a lower bound on the mass-to-light ratio in
the center of this cluster which is considerably larger than values
typically found in clusters by other means. 

The $\m$-statistics, which measures the projected density field
filtered with a compensated filter function, combines the
properties that it is directly related to the projected mass density,
and that it 
is obtainable from the shear, i.e., observable through image
ellipticities. Furthermore, in contrast to the mean 
shear within a circle, which is a two-component quantity from which no
non-trivial third-order moment can be defined, the $\m$-statistics is a
scalar whose skewness is well-defined.  Schneider (1996) has
suggested that the $\m$-statistics can be used to search for (dark)
matter concentrations on high-quality wide-field images.

In this paper we shall investigate the $\m$-statistics as a measure for
cosmic shear. In Sect.\ts 2 we briefly review the basic equations for
the light propagation in an inhomogeneous Universe, thereby
introducing our notation. The two-point $\m$-statistics in introduced in
Sect.\ts 3. It is shown that the dispersion of $\m$ on a certain
angular scale can be expressed as
an integral over the power spectrum of the projected surface mass
density, times a filter function containing this scale. This filter
function is shown to be quite narrow, so that $\ave{\m^2(\theta)}$
provides a fairly localized estimate of the power spectrum. This is
contrasted with the rms shear averaged on circles, for which the
corresponding filter function is very broad.

For various cosmological
models, the dispersion of $\ave{\m^2}$ is calculated, for both the
linear and fully non-linear growth of density perturbations, and
compared to the rms mean shear. In Sect.\ts 4, we calculate the
skewness of $\m$, following closely the treatment of BvWM, i.e.,
applying the quasi-linear theory of structure growth. 
Sect.\ts 5 is devoted to practical estimators of $\ave{\m^2}$
and the skewness, and their respective accuracies. A practical advantage of
the $\m$-statistics over the rms shear is that values of $\m$ measured in
neighboring apertures are virtually independent, whereas the
mean shear inside a circle has a very large correlation length. This
fact is of particular relevance for measurements of cosmic shear on
small angular scales, using wide-field cameras. For a fixed total
solid angle of available data, the signal-to-noise ratio for the
measurement of $\ave{\m^2}$ is obtained for $\theta\sim 2'$, somewhat
dependent on the kurtosis of the density field. Depending on
cosmology, a measurement of the dispersion of $\m$ with an accuracy of
$\sim 20$\% should be possible from one square-degree of deep imaging data.
We discuss our results in Sect.\ts 6. In the Appendix, we consider
several approximations which are used throughout the main body of the
paper. We focus on the contributions to the skewness which arise even
in the case of Gaussian density fluctuations, and
find that by including these effects, the skewness changes by 
$\sim 10$\%, which justifies the use of our approximations.

\sec{2 Light propagation in slightly inhomogeneous Universes}
We shall use a notation similar to K96 and JS. The metric of the
homogeneous Universe is written in the form
$$
\d s^2=c^2\,\d t^2 -a^2(t)\eck{\d w^2+f_K^2(w) \d\omega^2}\;,
\eqno (2.1)
$$
where $a(t)=(1+z)^{-1}$ is the dimensionless cosmic scale factor,
normalized to unity today; $w$ is a radial coordinate, and $f_K(w)$
is the {\it comoving angular diameter distance} to distance
$w$. The spatial curvature 
$$
K=\rund{H_0\over c}^2 \rund{\Omega_{\rm d}+\Omega_{\rm v} -1}\;,
\eqno (2.2)
$$
is related to the present day value of the density parameters
$\Omega_{\rm d}$ and $\Omega_{\rm v}$ in dust and in vacuum energy, in
terms of which $f_K(w)$ reads
$$
f_K(w)=\cases{ K^{-1/2} \sin\rund{\sqrt{K} w} & for $K>0$  , \cr
		w				& for $K=0$  , \cr
	(-K)^{-1/2} \sinh\rund{\sqrt{-K} w} & for $K<0$ . \cr}
\eqno (2.3)
$$
The angular diameter distance $D(z_1,z_2)$ is conveniently expressed
in terms of $f_K$ by
$$
D(z_1,z_2)={1\over 1+z_2} f_K(w_2-w_1)\;,
\eqno (2.4)
$$
$z_1\le z_2$, where the $w_i$ are the radial distances corresponding to redshifts
$z_i$,
$$
w=\int_0^z \d z' {c\over H}={c\over H_0}\int_0^z
{\d z'\over \sqrt{(1+z')^3 \Omega_{\rm d}+(1+z')^2 \rund{1-\Omega_{\rm
d} -\Omega_{\rm v}} +\Omega_{\rm v} } }\; .
\eqno (2.5)
$$
Light propagation in a weakly inhomogeneous Universe has been
investigated in many papers (e.g., Blandford et al.\ 1991; Seitz,
Schneider \& Ehlers 1994, and references therein). We shall use the
results of these papers in the following form: 

Consider a bundle of light rays intersecting at the observer. Each of
these rays is characterized by the angle $\vc\theta$ it encloses with
a fiducial ray. Let $\vc x(\vc\theta,w)$
denote the comoving transverse separation of the ray characterized by
$\vc\theta$ from the fiducial ray; it satisfies the propagation
equation 
$$
\dder{\vc x}{w}+K\vc x=-{2\over c^2}\eck{\nabla_\perp \Phi\rund{\vc
x\rund{\vc\theta,w},w}-\nabla_\perp \Phi^{(0)}(w)}\; ,
\eqno (2.6)
$$
where $\Phi\rund{\vc x\rund{\vc\theta,w},w}$ is the Newtonian
gravitational potential at comoving distance $w$ and comoving
perpendicular separation $\vc x\rund{\vc\theta,w}$ from the fiducial
ray, $ \Phi^{(0)}(w)$ is the potential along the fiducial ray, and
$\nabla_\perp=\rund{\part{ }{x_1},\part{ }{x_2}}$ is the transverse
gradient operator {\it in comoving coordinates}. Here we 
have assumed that the fiducial ray propagates nearly parallel to the
local $x_3$-direction, and since all angles involved are small, all
rays considered propagate nearly parallel to this direction. If the
Newtonian potential vanishes, $\vc x(\vc\theta,w)=f_K(w)\vc\theta$, in
agreement with the identification of $f_K(w)$ as comoving angular
diameter distance. A formal solution to (2.6) is obtained from the
Greens function of the operator on the left-hand side of (2.6),
yielding
$$
\vc x(\vc\theta,w)=f_K(w)\vc\theta-{2\over c^2}\int_0^w
\d w'\;f_K(w-w')\,\eck{\nabla_\perp \Phi\rund{\vc
x\rund{\vc\theta,w'},w'}-\nabla_\perp \Phi^{(0)}(w')} \;.
\eqno (2.7)
$$
A source at $w$ with comoving distance $\vc x$ from the fiducial ray
will be seen in the absence of light deflection at an angle
$\vc \beta=\vc x/f_K(w)$, which we shall call the unlensed source
position. Defining, as in usual lens theory, the Jacobian matrix $\A$ as
$$
\A(\vc\theta,w)={\partial \vc \beta\over \partial \vc\theta}
={1\over f_K(w)}{\partial \vc x\over \partial \vc\theta} \;,
\eqno (2.8)
$$
we obtain from differentiation of (2.7),
$$
\A_{ij}(\vc\theta,w)=\delta_{ij}-{2\over c^2}\int_0^w
\d w'\;{f_K(w-w')\,f_K(w')\over f_K(w)} \,\Phi_{,ik}\rund{\vc
x\rund{\vc\theta,w'},w'}\,\A_{kj}(\vc\theta,w')\; ,
\eqno (2.9)
$$
where indices $i$ on $\Phi$ preceded by a comma denote partial
derivatives w.r.t. $x_i$.
In general this is not an explicit equation for $\A$, since in
order to calculate $\A$, one first has to solve for the ray position
$\vc x(\vc\theta,w)$ and then solve the integral equation (2.9) for
$\A$. However, for weak gravitational fields which are of interest to
us here, one can expand $\A$ in powers of the Newtonian potential
$\Phi$, and keep only the lowest order term; this results in
$$
\A_{ij}(\vc\theta,w)=\delta_{ij}-{2\over c^2}\int_0^w
\d w'\;{f_K(w-w')\,f_K(w')\over f_K(w)}
\,\Phi_{,ij}\rund{f_K(w')\vc\theta,w'} \; .
\eqno (2.10)
$$
Hence, to linear order in $\Phi$, the distortion is obtained by
integrating the second derivatives of the potential along the
unperturbed ray. This feature allows for a dramatic simplification of the
calculations below. In the Appendix, we shall consider some
higher-order terms for $\A$ corresponding to lens-lens coupling and to
dropping the `Born approximation'; i.e., we shall calculate $\A$ up to
second order in $\Phi$. 
Also note that $\A$ given in (2.10) is
symmetric. One can therefore define a deflection potential,
$$
\psi(\vc\theta,w)={2\over c^2}\int_0^w
\d w'\;{f_K(w-w') \over
f_K(w')\,f_K(w)}\Phi\rund{f_K(w')\vc\theta,w'}\;,
\eqno (2.11)
$$
in terms of which one can treat lensing by large-scale structures
similarly to the single lens-plane case; e.g., the corresponding
`surface mass density' $\kappa(\vc\theta,w)$ and shear
$\gamma(\vc\theta)=\gamma_1+ {\rm i}\gamma_2$ are given as
$$\eqalign{
\kappa(\vc\theta,w)&={1\over 2}\rund{\psi_{,11}+\psi_{,22}}\; ,\cr
\gamma(\vc\theta,w)&={1\over 2}\rund{\psi_{,11}-\psi_{,22}}+{\rm
i}\psi_{12}\; ,\cr }
\eqno (2.12)
$$
where indices on $\psi$ separated by a comma denote partial
derivatives with respect to $\theta_i$. In particular,
$\A_{ij}=\delta_{ij}-\psi_{,ij}$. Thus, one can consider the surface
mass density $\kappa$ -- or its associated deflection potential $\psi$
-- as the fundamental quantity, and in fact it is
the only quantity which is probed by cosmic shear measurements. One
obtains from (2.11) and (2.12), by adding
a term $\Phi_{,33}$ to the integrand which cancels out upon $w$-integration, 
and by using Poisson's equation in the form
$$
\nabla^2\Phi={3 H_0^2 \Omega_{\rm d}\over 2 a}\delta\;,
\eqno (2.13)
$$
the following expression for the projected density field:
$$
\kappa(\vc\theta,w)={3\over 2}\rund{H_0\over c}^2\Omega_{\rm d}
\int_0^w
\d w'\;{f_K(w-w')\,f_K(w')\over f_K(w)}
{\delta\rund{f_K(w')\vc\theta,w'}\over a(w')} \;,
\eqno (2.14)
$$
where $\delta(\vc x,w)$ is the density contrast. The projected density
field depends on the source redshift (or distance). We shall assume
that one observes the shear through a population of galaxies for which
only the redshift probability distribution $p_z(z)$ is known, or
equivalently, $p_w(w)\d w=p_z(z)\d z$. Then, the source
distance-averaged projected mass density becomes
$$
\kappa(\vc\theta):=\int \d w\; p_w(w)\,\kappa(\vc\theta,w) =
{3\over 2}\rund{H_0\over c}^2\Omega_{\rm d}
\int_0^{w_{\rm H}} \d w\;
g(w)\,f_K(w){\delta\rund{f_K(w)\vc\theta,w}\over a(w)}\; ,
\eqno (2.15)
$$
where
$$
g(w):=\int_w^{w_{\rm H}}\d w'\;p_w(w')\,{f_K(w'-w)\over f_K(w')}
\eqno (2.16)
$$
is the source-averaged distance ratio $D_{\rm ds}/D_{\rm s}$ for a
density fluctuation at distance $w$, and $w_{\rm H}$ is the comoving
distance to the horizon, obtained by putting $z=\infty$ in (2.5).
In this paper, we shall consider two different redshift distributions
of sources. In the first case, all sources are assumed to reside at the
same redshift $z_{\rm s}$, so that
$$
p_z(z)=\delta_{\rm D}(z-z_{\rm s})\;.
\eqno (2.17)
$$
More realistically, we consider a redshift distribution of the form
(e.g., Smail et al.\ts 1995)
$$
p_z(z)\propto z^2\,\exp\rund{-[z/z_0]^\beta}\;.
\eqno (2.18)
$$
The mean redshift of this distribution is proportional to $z_0$ and
depends on the parameter $\beta$ which describes how quickly the
distribution falls off towards higher redshifts. In particular, for
$\beta =1.5$ (a value that we shall use throughout), $\ave{z}=1.505
z_0$.

\sec{3 Two-point $\m$-statistics}
\subs{3.1 The power spectrum of the projected density field}
Provided the density contrast $\delta$ is a homogeneous and isotropic
random field, so is the projected density $\kappa$. 
Consider the Fourier transform of the projected density field,
$$
\tilde\kappa(\vc s):=\int \d^2\theta\;\kappa(\vc\theta)\,{\rm
e}^{-{\rm i}\vc\theta\cdot\vc s}\;.
\eqno (3.1)
$$
We define the power spectrum $P_\kappa(s)$ of $\kappa$ by
$$
\ave{\tilde\kappa(\vc s) \tilde\kappa^*(\vc s')}
=(2\pi)^2 \delta_{\rm D}(\vc s-\vc s')\,P_\kappa(\abs{\vc s})\;,
\eqno (3.2)
$$
where the Dirac delta `function' $\delta_{\rm D}$ occurs because
$\kappa(\vc\theta)$ is a homogeneous random field, and the dependence
of $P_\kappa$ on the modulus of $\vc s$ only expresses the fact that
$\kappa$ is an isotropic random field. Following K96, we can calculate
$P_\kappa(s)$ from the power spectrum of the density fluctuations,
defined accordingly by
$$
\ave{\tilde\delta(\vec k)\tilde\delta^*(\vec k')}=
(2\pi)^3 \delta_{\rm D}(\vec k-\vec k')\,P(\vert \vec k\vert)\; .
\eqno (3.3)
$$
The derivation of the Fourier-space analog of Limber's equation in K96
is valid provided the power spectrum $P$ does not evolve appreciably
on timescales corresponding to the light travel time across the
largest significant fluctuations, and provided the typical source
distance is much larger than the largest-scale fluctuations. With
these two assumptions, K96 (see also Kaiser 1992) obtains
$$
P_\kappa(s) =
{9\over 4}\rund{H_0\over c}^4\Omega_{\rm d}^2
\int_0^{w_{\rm H}}\d w\;{g^2(w)\over a^2(w)} 
P\rund{{s\over f_K(w)};w}\; .
\eqno (3.4)
$$
The second argument of $P$ indicates that the power spectrum 
evolves (slowly) with redshift. We shall later derive an analogous
relation for the three-point function, using the same strategy as in
the derivation of (3.4). Since $s$ is the Fourier-conjugate
of the angle $\theta$, we can relate an angular scale to its
corresponding $s$ by $s=2\pi/\theta=2.16\times 10^4
\,(\theta/1\ts{\rm arcmin})^{-1}$. 

In Fig.\ts 1, we have plotted $P_\kappa(s)$ for five different
cosmological models. For three of them, the 
power spectrum $P(k)$ is approximately cluster normalized, which
corresponds to $\sigma_8\approx 0.6$ for an Einstein-de Sitter
universe (EdS, $\Omega_{\rm d}=1$, $\Omega_{\rm v}=0$), $\sigma_8=1$
for both an open universe (OCDM, $\Omega_{\rm d}=0.3$,
$\Omega_{\rm v}=0$) and a spatially flat universe with cosmological constant
($\Lambda$CDM, $\Omega_{\rm d}=0.3$, $\Omega_{\rm v}=0.7$). In all
these cases, we have used the CDM spectrum as given by Bardeen et
al.\ts (1986), but set the shape parameter of the linear power
spectrum to $\Gamma=0.25$, which yields the best fit to the observed
two-point correlation function of galaxies (Efstathiou 1996). The
remaining two cosmological models have EdS geometry, but either a
higher normalization ($\sigma_8=1$, approximately corresponding to the
COBE normalization) or a different shape parameter ($\Gamma=0.5$,
corresponding to the original definition for $\Gamma$ if $H_0=50$\ts
km/s/Mpc). For each model, the projected power spectrum has been
calculated for a linearly-evolved cosmological power-spectrum (thin
curves), and for the fully non-linear power spectrum, following
the prescriptions of Hamilton et al.\ts (1991), Jain, Mo \& White
(1995), and Peacock \& Dodds (1996) -- we used the fit formulae of the
latter paper throughout. 

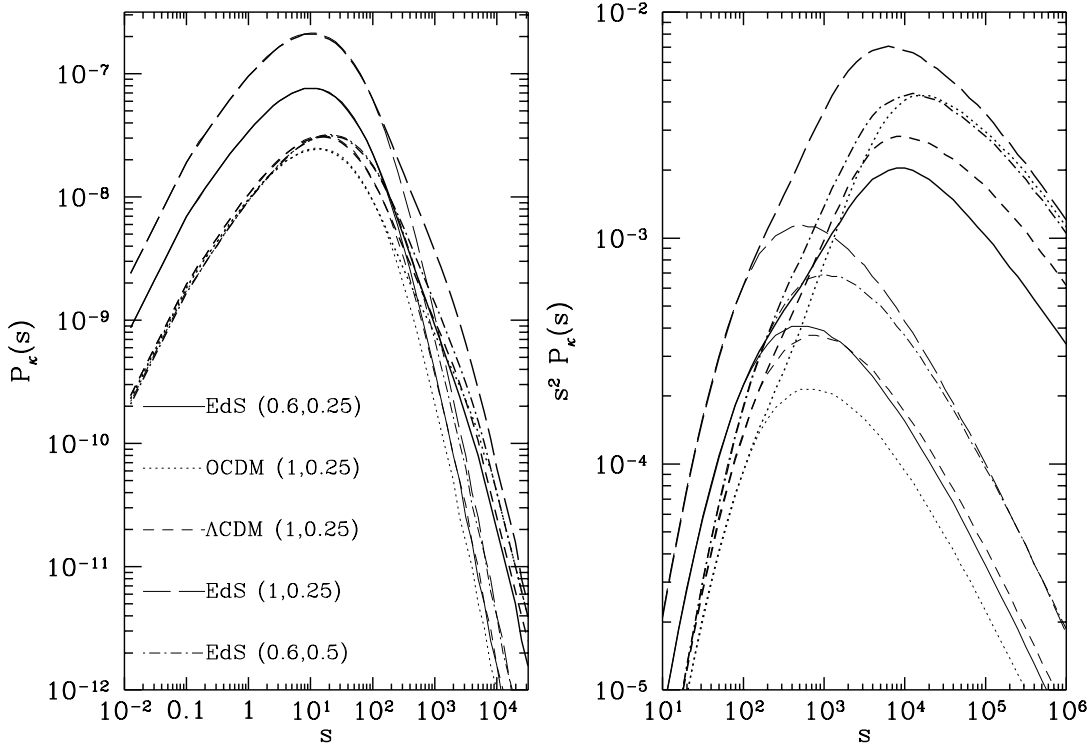
\xfigure{1}{The power spectrum $P_\kappa(s)$ of the projected density
field, as defined in (3.4), for five different cosmological models, as
indicated by the line types; the numbers in
parenthesis are $(\sigma_8,\Gamma)$.
For each cosmological model, the thin curves
correspond to the linear evolution of the power spectrum $P(k,a)$,
whereas the thick curves are calculated with the fully non-linear
evolution of the power spectrum, as given in Peacock \& Dodds
(1996). For this plot, the redshift distribution (2.18) was used, with
$z_0=1$ and $\beta=1.5$. Note that the large amplitude of the EdS
model with $\gamma=0.25$ relative to the OCDM and $\Lambda$CDM model
is due to the factor $\Omega_{\rm d}^2$ in (3.4)
}{fig1.tps}{15}

\subs{3.2 The $\m$-statistics}
Define the aperture mass by
$$
\m(\theta):=\int \d^2\vt\;U\rund{\abs{\vc\vt}}
\kappa(\vc\vt)\;,
\eqno (3.5)
$$
where the integral extends over a circle of angular radius $\theta$, and
$U(\vt)$ is a continuous weight function which vanishes for $\vt>\theta$.
Provided $U$ is a compensated filter function, i.e.,
$$
\int_0^\theta \d\vt\;\vt \, U(\vt)=0\;,
\eqno (3.6)
$$
one can express $\m$ in terms of the tangential component $\gamma_{\rm
t}$ of the shear
inside the circle (Kaiser et al.\ 1994; Schneider 1996),
$$
\m(\theta)=\int \d^2\vt\;Q\rund{\abs{\vc\vt}}\,\gamma_{\rm
t}(\vc\vt)\; ,
\eqno (3.7)
$$
where
$$
Q(\vt)={2\over \vt^2}\int_0^\vt\d \vt'\; \vt'\,U(\vt')
-U(\vt) \;,
\eqno (3.8)
$$
and the tangential component of the shear at a position
$\vc\vt=(\vt \cos\vp,\vt\sin\vp)$ is
$$
\gamma_{\rm t}(\vc\vt)=-\Re \rund{\gamma(\vc\vt)\, {\rm
e}^{-2{\rm i}\vp}}\; .
\eqno (3.9)
$$
The useful property of $\m$ is thus that, one the one hand, it yields a
spatially filtered version of the projected density field, and on the
other hand, that it can be expressed simply in terms of the shear. Since in
the weak lensing regime, the observed galaxy ellipticities provide an
unbiased estimate of the local shear, $\m$ is directly related to
observables. 

Obviously, the ensemble average of $\m$ vanishes, $\ave{\m}=0$. The
dispersion of $\m$ can be calculated as follows:
$$\eqalign{
\ave{\m^2(\theta)}&=\int\d^2\theta'\;U(\theta')\int\d^2\vt\;U(\vt)
\ave{\kappa(\vc \theta')\kappa(\vc\vt)} \cr
&=\int\d^2\theta'\;U(\theta')\int\d^2\vt\;U(\vt)
\int{\d^2 s\over (2\pi)^2}\;{\rm e}^{{\rm i}\vc
s\cdot(\vc\theta'-\vc\vt)} P_\kappa(s) \cr
&=2\pi \int_0^\infty \d s\;s\,P_\kappa(s)
\rund{\int_0^\theta\d\vt\;\vt\,U(\vt)\,{\rm J}_0(s\vt)}^2\;
,\cr }
\eqno (3.10)
$$
where we used in the first step that the two-point correlation
function is the Fourier transform of the power spectrum, and the
Bessel function ${\rm J}_0$ comes from the angular integrations in
$\vc\theta'$ and $\vc\vt$. 

The choice of the weight function $U(\vt)$ is arbitrary at this
point. We shall write $U(\vt)=u(\vt/\theta)/\theta^2$, with $u(x)=0$ 
for $x>1$, and $Q(\vt)=q(\vt/\theta)/\theta^2$.
Furthermore, we choose the normalization such that
$$
2\pi \int_0^\theta\d\vt\; \vt\,Q(\vt)
=2\pi \int_0^1\d x\;x\,q(x) =1\; .
\eqno (3.11)
$$
A set of weight functions which satisfy (3.6) and (3.11) is
$$
u(x)={(\ell+2)^2\over \pi}\rund{1-x^2}^\ell\rund{{1\over \ell+2}-x^2}\;,
\eqno (3.12)
$$
which peak at $x=0$ and go to zero with order $\ell$ as $x\to 1$. 
Correspondingly,
$$
q(x)={(1+\ell)(2+\ell)\over \pi} x^2 (1-x^2)^\ell\; .
\eqno (3.13)
$$
Then defining
$$\eqalign{
I_\ell(\eta)&:={(\ell+2)^2\over \pi}\int_0^1\d x\;x\,\rund{1-x^2}^\ell
\rund{{1\over \ell+2}-x^2} {\rm J}_0(\eta x) \cr
&= {2^\ell\Gamma(\ell +3)\over \pi} \eta^{-(\ell+1)}\,{\rm
J}_{3+\ell}(\eta) \;, \cr }
\eqno (3.14)
$$
we can write the dispersion as
$$
\ave{\m^2(\theta)}=2\pi\int_0^\infty \d s\;s\,P_\kappa(s)\,
\eckk{I_\ell(s\theta)}^2\; .
\eqno (3.15)
$$
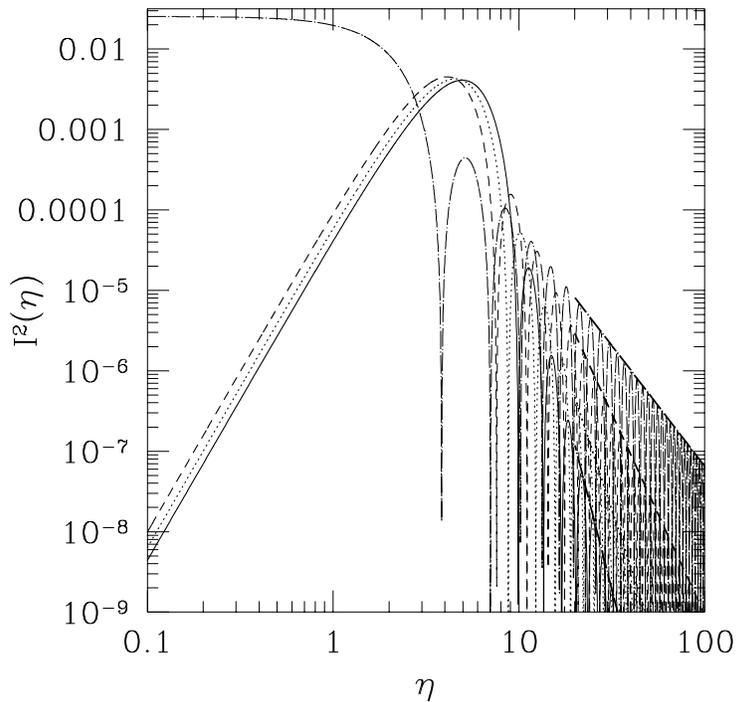
\xfigure{2}{The filter function $I_\ell^2(\eta)$ in the dispersion of $\m$
for three different value of the $\ell$: $\ell=1$ (dashed curve),
$\ell=2$ (dotted curve), and $\ell=3$ (solid curve). For comparison,
the corresponding filter function for top-hat filtering $I^2_{\rm
TH}(\eta)$ is also plotted (dashed-dotted curve). For larger values of
$\eta$, all functions rapidly oscillate, owing to the Bessel functions.
For clarity, we have therefore plotted the amplitude of these
oscillations as corresponding thick curves}{fig2.tps}{10}
We have plotted the filter function $I_\ell^2(\eta)$ in Fig.\ts 2, for
three different values of $\ell$. One can see from (3.14) that
$I_\ell(\eta)\propto \eta^2$ for small $\eta$, and that for large $\eta$,
the function $I_\ell(\eta)$ oscillates with an amplitude $\propto
\eta^{-(\ell+3/2)}$. Hence, the filter function $I_\ell^2$ behaves like
$\eta^4$ and $\eta^{-(2\ell+3)}$ in the two respective limits, and is
indeed a very localized filter, with a peak at $\eta\sim 5$ for small
$\ell$. Hence, $\ave{\m^2(\theta)}$ provides an accurate measure of the
power spectrum of the projected density field at $s\sim 5/\theta$, with
little dependence on the local shape of this power spectrum. 

We shall compare $\ave{\m^2(\theta)}$ with the rms value of the shear
averaged over a circular region of angular radius $\theta$: let 
$$
\bar\gamma(\theta):={1\over\pi \theta^2}\int\d^2\vt\; \gamma(\vc\vt)\;,
\eqno (3.16)
$$
then its dispersion is
$$
\ave{\abs{\bar\gamma}^2(\theta)}=2\pi\int_0^\infty \d s\;s\,
P_\kappa(s)\,\eckk{I_{\rm TH}(s\theta)}^2\; ,
\eqno (3.17)
$$
with
$$
I_{\rm TH}(\eta)={{\rm J}_1(\eta)\over \pi \eta}\; 
\eqno (3.18)
$$
(cf. Blandford et al.\ 1991). The dash-dotted curve in Fig.\ts 2
displays $I_{\rm TH}^2$ for comparison with $I_\ell$. One sees
that $I_{\rm TH}^2$ is a much broader function which tends towards a
constant for $\eta\to 0$, and its amplitude decreases like $\eta^{-3}$
for large $\eta$. Therefore, the shear dispersion
$\ave{\abs{\bar\gamma}^2}$ is a much coarser probe of the power spectrum
than $\ave{\m^2(\theta)}$. 


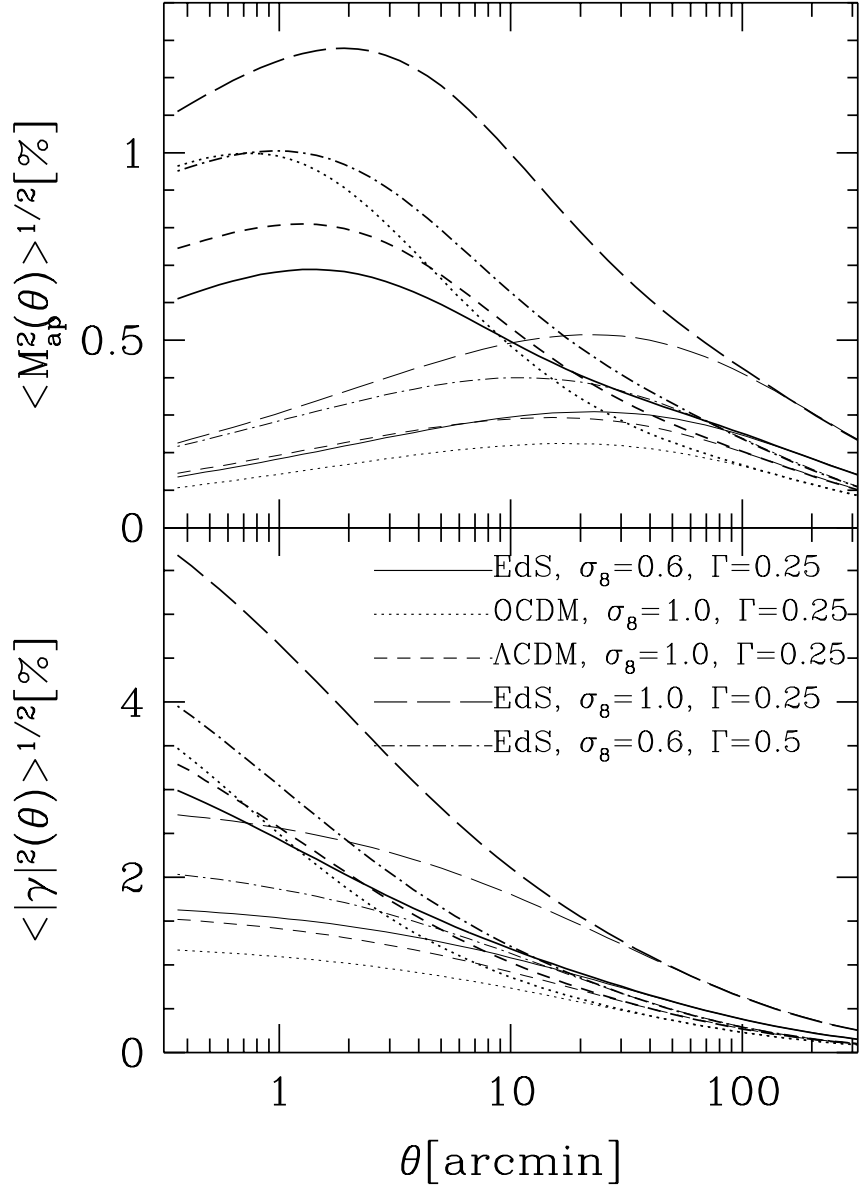
\xfigure{3}{The rms values of the shear are shown vs. filter scale $\theta$
for the same cosmological models as used in Fig.\ts 1. The upper panel
shows the rms value 
of $\m$ with the filter $I_1$ defined in (3.14), while the lower
panel shows the rms shear computed with a top-hat filter (3.18). Note
the different scales in the two panels. The 
source galaxies are assumed to follow the distribution (2.18), with
$\beta=1.5$ and $z_0=1$. The thin curves display the prediction for
the rms shear if linear evolution of the density fluctuation spectrum
is assumed, whereas the thick curves follow from the fully non-linear
evolution of the power spectrum. The results shown in the lower panel are
fully equivalent to those of JS, except that they considered a
redshift distribution of the form (2.17), and they considered the
`polarization' as a measure of
net galaxy ellipticity, which in the weak lensing case equals twice
the shear}{fig3.tps}{13}

\subs{3.3 Results}
In Fig.\ts 3 we have plotted the rms values of the $\m$-statistics
(upper panel), as well as that of the mean shear within circles (lower
panel). The same cosmological models as in Fig.\ts 1 were used, and we
present results both for the linearly evolved cosmic power spectrum
(thin curves) and for the fully non-linear evolution (thick
curves). Whereas $\ave{\abs{\bar\gamma}^2}^{1/2}$ decreases monotonically
with $\theta$, the shape of $\ave{\m^2}^{1/2}$ closely reflects the
shape of the projected power spectrum $P_\kappa$ displayed in Fig.\ts
1; this is of course related to the narrowness of the filter $I_\ell$
shown in Fig.\ts 2. In Fig.\ts 3, and for the figures of the
remainder of the paper, we used $\ell=1$, but have checked that
changing to $\ell=2$ or 3 does not yield qualitatively different
results. Compared to the prediction of the linear evolution of the
power spectrum, the peak of $\ave{\m^2}^{1/2}$ is shifted to
substantially smaller angles; at the same time, the non-linear
evolution affects $\ave{\m^2}^{1/2}$ more than
$\ave{\abs{\bar\gamma}^2}^{1/2}$, as the latter picks up power from larger
scales which are hardly affected by non-linear evolution.

As can be seen from the figure, the values of $\ave{\m^2}^{1/2}$ are
substantially smaller than those of $\ave{\abs{\bar\gamma}^2}^{1/2}$, at
least on scales below one degree. This is related to the fact that at
a given angular scale $\theta$, the $\m$-statistics is sensitive to
smaller-scale structures than the $\abs{\bar\gamma}$-statistics, as can be
seen in Fig.\ts 2, and so these two rms values should be compared at
different `effective' scales. Whereas this difference in magnitude may
suggest that the $\m$-statistics is observationally disfavoured, 
we shall show in Sect.\ts 5 that it has the advantage that
measurements in neighboring fields are nearly uncorrelated. 

The dependence of the rms of $\m$ on the source redshift distribution
is displayed in Fig.\ts 4, for the same five cosmological models, and
two angular scales. The dependence
on $z_{\rm s}$ is generally weak, in rough agreement with the 
power law dependence $z_{\rm s}^{0.6}$ found by JS for the top-hat filter. 
The model with cosmological constant shows a stronger growth with
$z_{\rm s}$, as distances grow faster with redshift in such models, and
thus provide larger path lengths for lensing. 

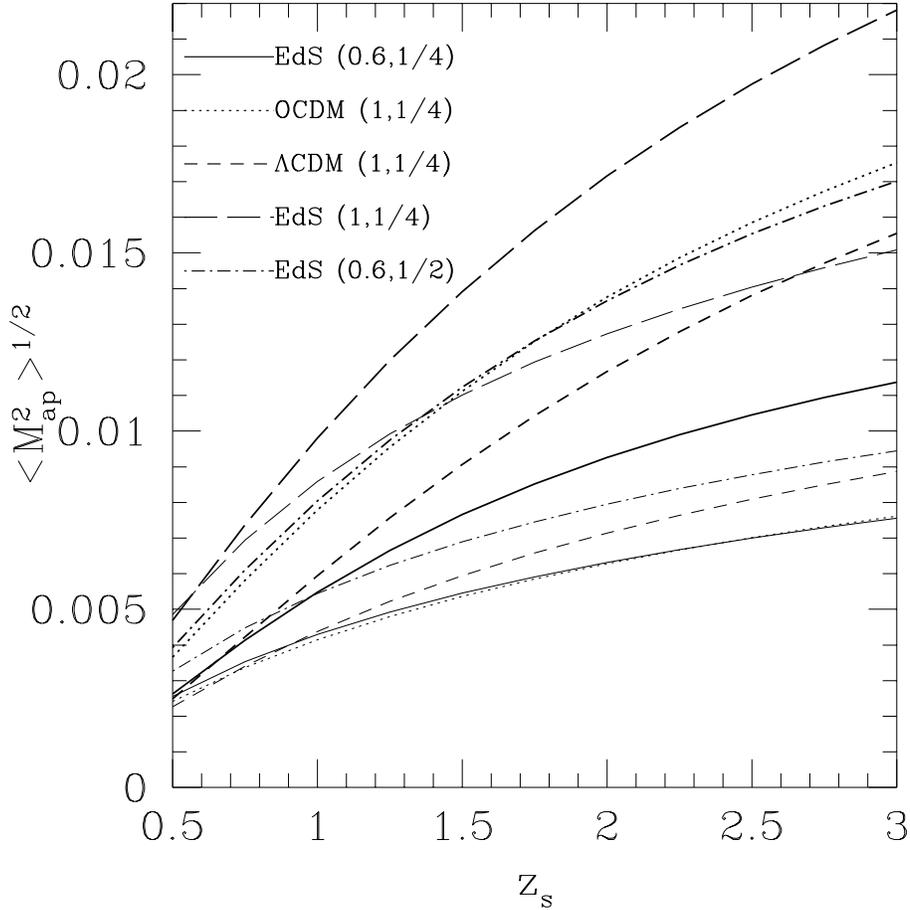
\xfigure{4}{The rms value of $\m$ as a function of source
redshift, for the same five cosmological models as in Fig.\ts 1,
indicated in the figure; the numbers in parenthesis are
$(\sigma_8,\Gamma)$. The
thick curves are for $\theta=1'$, the thin curves for
$\theta=10'$. The fully non-linear evolution of the power spectrum has
been used. The sources were assumed to be all at the same redshift
$z_{\rm s}$}{fig4.tps}{13}

\def\tk{{\tilde\kappa}}
\sec{4 Three-point $\m$-statistics: skewness}
If $\kappa$ is a Gaussian random field, so is $\m$,
and in particular, the expectation value of the third-order moment of
$\m$ would vanish. However, since the nonlinear gravitational 
evolution of density
fluctuations transforms an initially Gaussian field into a
non-Gaussian one, this third-order moment is non-zero in general. In
fact, $\ave{\m^3}$ is a measure of the non-Gaussianity of the density
fluctuations at medium redshifts. Since
$$\eqalign{
\ave{\m^3(\theta)}&=\int\d^2\theta_1\;U(\theta_1)
\int\d^2\theta_2\;U(\theta_2)
\int\d^2\theta_3\;U(\theta_3)
\ave{\kappa(\vc\theta_1)\kappa(\vc\theta_2)\kappa(\vc\theta_3)} \cr
&=\int\d^2\theta_1\;U(\theta_1)
\int\d^2\theta_2\;U(\theta_2)
\int\d^2\theta_3\;U(\theta_3) \cr &\times
\int{\d^2 s_1\over (2\pi)^2} {\rm e}^{{\rm i}\vc \theta_1\cdot \vc s_1}
\int{\d^2 s_2\over (2\pi)^2} {\rm e}^{{\rm i}\vc \theta_2\cdot \vc s_2}
\int{\d^2 s_3\over (2\pi)^2} {\rm e}^{{\rm i}\vc \theta_3\cdot \vc s_3}
\ave{\tk(\vc s_1) \tk(\vc s_2) \tk(\vc s_3)}\; ,\cr }
\eqno (4.1)
$$
the evaluation of the third moment requires the calculation of the
three-point correlation function of $\tk$. 

The skewness of the projected surface mass density has already been
discussed in BvWM. They have calculated the skewness of the projected
density field by considering
the quasi-linear evolution of the density fluctuations. Below, we shall
follow the same procedure to calculate $\ave{\m^3}$. As was also pointed
out by BvWM, even for the strictly linear evolution of the power
spectrum, when the density field conserves its initial Gaussian
nature, the observable skewness would not vanish identically.
We get back to this point at the end of this section.

\subs{4.1 The three-point correlator for $\tk$}
We shall now evaluate this three-point function in terms of the
corresponding function of the cosmic density field, i.e., to obtain
the analog of (3.4) for the three-point function. For abbreviation, we
write (2.14) in the form
$$
\kappa(\vc\theta,w)=\int_0^w \d w'\;G(w,w')\,\delta\rund{f_K(w')\vc\theta,w'}\;,
\eqno (4.2)
$$
with
$$
G(w,w')={3\over 2}\rund{H_0\over c}^2\Omega_{\rm d}{f_K(w-w')\,
f_K(w')\over f_K(w)\,a(w')}
$$
for $w'\le w$, and zero otherwise. Then,
$$\eqalign{
&\ave{\tk(\vc s_1,w_1)\tk(\vc s_2,w_2)\tk(\vc s_3,w_3)}\cr
&=\int\d^2\theta_1\;{\rm e}^{-{\rm i}\vc s_1\cdot \vc\theta_1}
\int\d^2\theta_2\;{\rm e}^{-{\rm i}\vc s_2\cdot \vc\theta_2}
\int\d^2\theta_3\;{\rm e}^{-{\rm i}\vc s_3\cdot \vc\theta_3}\cr
&\times
\int_0^{w_1}\d v_1\; G(w_1,v_1) \int_0^{w_2}\d v_2\;G(w_2,v_2)
\int_0^{w_3} \d v_3\;G(w_3,v_3) \cr 
&\times
\ave{\delta\rund{f_K(v_1)\vc\theta_1,v_1} 
\delta\rund{f_K(v_2)\vc\theta_2,v_2} 
\delta\rund{f_K(v_3)\vc\theta_3,v_3} } \; , \cr }
\eqno (4.3)
$$
where we have inserted the Fourier transforms of $\tk$ and used
(4.2). Assuming that the largest scale fluctuations are much smaller
than the typical distance to a source, the three-point correlation
function of $\delta$ will vanish except when $v_1\approx v_2\approx
v_3$. Thus over the $v$-range where $\ave{\delta\delta\delta}$ does
not vanish, we can set the $v$-arguments in $G$ to be all the same,
and we can also set all $f_K(v_i)$ equal. Since $G(w,v)=0$ if $v>w$,
the outer $v$-integration only extends to $w_{\rm
min}=\min(w_1,w_2,w_3)$. Hence, after replacing $\delta$ by its
Fourier transform, one obtains
$$
\eqalign{
&\ave{\tk(\vc s_1,w_1)\tk(\vc s_2,w_2)\tk(\vc s_3,w_3)}\cr
&=\int\d^2\theta_1\;{\rm e}^{-{\rm i}\vc s_1\cdot \vc\theta_1}
\int\d^2\theta_2\;{\rm e}^{-{\rm i}\vc s_2\cdot \vc\theta_2}
\int\d^2\theta_3\;{\rm e}^{-{\rm i}\vc s_3\cdot \vc\theta_3}\cr
&\times
\int_0^{w_{\rm min}}\d v\;G(w_1,v) G(w_2,v) G(w_3,v)
\int \d v' \int \d v'' \cr
&\times
\int{\d^3 k_1\over (2\pi)^3}\; {\rm e}^{{\rm i}f_K(v)\vc
\theta_1\cdot\vc k_1} 
{\rm e}^{{\rm i}k_{13} v}
\int{\d^3 k_2\over (2\pi)^3}\; {\rm e}^{{\rm i}f_K(v)\vc
\theta_2\cdot\vc k_2}
{\rm e}^{{\rm i}k_{23} v'} \cr
&\times 
\int{\d^3 k_3\over (2\pi)^3}\; {\rm e}^{{\rm i}f_K(v)\vc
\theta_3\cdot\vc k_3}
{\rm e}^{{\rm i}k_{33} v''} 
\ave{\tilde\delta(\vec k_1) \tilde\delta(\vec
k_2) \tilde \delta(\vec k_3)}
 \; ,}
\eqno (4.4)
$$
where we have written the 3-dimensional vector $\vec k$ as $(\vc
k,k_3)$. The $\theta_i$-integrations can now be carried out, each
yielding $(2\pi)^2 \delta_{\rm D}(\vc s_i-f_K(v) \vc k_i)$. After
that, the $\vc k_i$-integrations become trivial. 
The $v'$
and $v''$-integrations can be carried out yielding delta-functions in
$k_{23}$ and $k_{33}$, which are trivially integrated away. One thus
finds:
$$\eqalign{
&\ave{\tk(\vc s_1,w_1)\tk(\vc s_2,w_2)\tk(\vc s_3,w_3)} 
=\int_0^{w_{\rm min}}\d v\;G(w_1,v) G(w_2,v) G(w_3,v){1\over f_K^6(v)} \cr
&\times
\int{\d k_3\over (2\pi)}\;
{\rm e}^{{\rm i}k_{3} v}
\ave{\tilde\delta\rund{{\vc s_1\over f_K(v)},k_3}
\tilde\delta\rund{{\vc s_2\over f_K(v)},0}
\tilde\delta\rund{{\vc s_3\over f_K(v)},0} }(v) \; . \cr }
\eqno (4.5)
$$
Next, the average of (4.5) over a source redshift distribution is
performed. The only point to notice here is the upper limit of
integration for $v$. Integrating (4.5) over $\prod \int \d w_i\,p_w(w_i)$
(recall that $p_w(w)\,\d w=p_z(z)\,\d z$ is the source redshift distribution)
gives an expression of the form
$$\eqalign{
I&=\int_0^{w_{\rm H}}\d w_1\int_0^{w_{\rm H}}\d w_2\int_0^{w_{\rm H}}\d
w_3 \int_0^{w_{\rm min}} \d v\; F\; \cr
&=\int_0^{w_{\rm H}}\d v\int_v^{w_{\rm H}}\d w_1\int_v^{w_{\rm H}} \d w_2
\int_v^{w_{\rm H}}\d w_3\; F \; . \cr
}
$$
In obtaining the second equality we have made use of the fact
that the integrand $F$ is symmetric in $w_1,w_2,w_3$, and that
$G(w,v)=0$ if $v>w$. 
We therefore obtain for the
redshift-averaged three-point correlation
$$\eqalign{
\ave{\tk(\vc s_1) \tk(\vc s_2) \tk(\vc s_3)}
&={27\over 8}\rund{H_0\over c}^6\Omega_{\rm d}^3
\int_0^{w_{\rm H}}\d w\;{g^3(w)\over a^3(w)f_K^3(w)} \cr
\times
\int{\d k_3\over (2\pi)}&\;{\rm e}^{{\rm i}k_{3} w}
\ave{\tilde\delta\rund{{\vc s_1\over f_K(w)},k_3}
\tilde\delta\rund{{\vc s_2\over f_K(w)},0}
\tilde\delta\rund{{\vc s_3\over f_K(w)},0} } \; , \cr }
\eqno (4.6)
$$
with $g(w)$ as defined in (2.16).

\def\td{{\tilde\delta}}
\subs{4.2 Quasi-linear theory of density fluctuations}
In order to calculate the triple correlator in (4.6), we shall use 
quasi-linear perturbation theory, in which the density field $\delta$
is considered as a `small' quantity and expanded into a perturbation
series, $\delta=\delta^{(1)}+\delta^{(2)}+\dots$, where 
$\delta^{(n)}=\O\rund{\eckk{\delta^{(1)}}^n}$. Here, $\delta^{(1)}$ is
the linearly evolved density perturbation, $\tilde\delta^{(1)}(\vec k,w)=
D_+(w)\tilde\delta^{(1)}_0(\vec k)$, $D_+(w)$ is the linear growth factor,
normalized to $D_+(0)=1$, and $\tilde\delta^{(1)}_0(\vec k)$ is the
density perturbation linearly extrapolated to the present epoch. The
perturbation series 
can then be inserted into the continuity equation and the Euler
equation, and a closed solution for every order can be obtained in
terms of lower-order terms (see, e.g., Fry 1984, Goroff et al.\ts 1986,
and references therein). In particular, in an EdS Universe, for the
first non-linear term, one obtains
$$\eqalign{
\td^{(2)}(\vec k,w)&=D_+^2(w)\int{\d^3 k'\over (2\pi)^3}\;\td^{(1)}_0(\vec k')\;
\td^{(1)}_0(\vec k-\vec k') \cr
&\times
\eck{{5\over 7}+{\vec k'\cdot \rund{\vec k-\vec k'} \over \abs{\vec k'}^2}
+{2\over 7}{\eck{\vec k' \cdot\rund{\vec k-\vec k'}}^2\over
\abs{\vec k'}^2 \abs{\vec k- \vec k'}^2}}\; . \cr }
\eqno (4.7)
$$
Bouchet et al.\ts (1992) showed that the $\vec k$-dependence of this
term depends weakly on cosmology, and that it is an excellent
approximation to restrict the cosmology dependence solely to
the growth factor $D_+^2(w)$.  In order to calculate $\ave{\td \td
\td}$, we note that the tri-linear correlation of the linear density
field vanishes, owing to its assumed Gaussian nature. Thus to lowest
order, we have
$$\eqalign{
\ave{\td(\vec k_1)\td(\vec k_2)\td(\vec k_3)}(w)
&=\ave{\td^{(1)}(\vec k_1,w)\td^{(1)}(\vec k_2,w)\td^{(2)}(\vec
k_3,w)} \cr
&+ \hbox{2 terms obtained from permutation}\; . \cr }
\eqno (4.8)
$$
Considering only the first term, inserting (4.7), making use of the
fact that $\td^{(1)}(\vec k,w)=D_+(w)\td^{(1)}_0(\vec k)$, and using
the relation
$$\eqalign{
&\ave{\td^{(1)}(\vec k_1)\td^{(1)}(\vec k_2)
\td^{(1)}(\vec k_3)\td^{(1)}(\vec k_4)} 
=\ave{\td^{(1)}(\vec k_1)\td^{(1)}(\vec k_2)} 
\ave{\td^{(1)}(\vec k_3)\td^{(1)}(\vec k_4)} \cr
&+ \hbox{2 terms obtained from permutation} \cr
&=(2\pi)^6 \rund{P_0(k_1)P_0(k_3)\delta_{\rm D}(\vec k_1+\vec k_2)
\delta_{\rm D}(\vec k_3+\vec k_4)} +\hbox{2 terms}\; , \cr }
\eqno (4.9)
$$
valid for Gaussian fields, we find
$$\eqalign{
&\ave{\td^{(1)}(\vec k_1)\td^{(1)}(\vec k_2)\td^{(2)}(\vec
k_3)}(w)
=2(2\pi)^3 D_+^4(w)P_0(k_1)P_0(k_2)\delta_{\rm D}(\vec k_1+\vec
k_2+\vec k_3) \cr
&\times
\eck{{5\over 7}+{1\over 2}\rund{{1\over\abs{\vec k_1}^2}+{1\over \abs{\vec
k_2}^2}} 
{\vec k_1\cdot\vec k_2} +{2\over 7}{\eck{\vec k_1\cdot\vec k_2}^2\over
\abs{\vec k_1}^2\abs{\vec k_2}^2}}\; . \cr }
\eqno (4.10)
$$
Here, $P_0(k)$ is the power spectrum of the linearly extrapolated
density field.
This expression is only one of three terms appearing in (4.8), with
the other two being obtained by permutation of the $\vec
k_i$. In order to calculate $\m^3$ according to (4.1), we have to
integrate over all $\vc s_i$, and each of the three terms in (4.8)
yields the same contribution. Therefore, we can just use (4.10) instead
of (4.8), and multiply the result by a factor 3. Then, combining
(4.1), (4.6) and (4.10), we obtain
$$\eqalign{
\ave{\m^3(\theta)}&={81\over 4}(2\pi)^{-1} \rund{H_0\over c}^6\Omega_{\rm d}^3
\int_0^{w_{\rm H}}\d w\;{g^3(w)\,D_+^4(w)\over a^3(w)\,f_K(w)} \cr
&\times
\int\d^2 s_1\;P_0\rund{s_1\over f_K(w)}\,I(s_1 \theta)
\int\d^2 s_2\;P_0\rund{s_2\over f_K(w)}\,I(s_2 \theta) \cr
&\times
I\rund{\abs{\vc s_1+\vc s_2}\theta}
\eck{{5\over 7}+{1\over 2}\rund{{1\over\abs{\vc s_1}^2}+{1\over \abs{\vc
s_2}^2}} 
{\vc s_1\cdot\vc s_2} +{2\over 7}{\eck{\vc s_1\cdot\vc s_2}^2\over
\abs{\vc s_1}^2\abs{\vc s_2}^2}} \; . \cr }
\eqno (4.11)
$$

\subs{4.3 Skewness}
We define the skewness as
$$
S(\theta):={\ave{\m^3(\theta)}\over \ave{\m^2(\theta)}^2}\;.
\eqno (4.12)
$$
We have calculated $S(\theta)$ using the results of the preceding
subsection. Quasi-linear theory is expected to underestimate
$\ave{\m^3(\theta)}$, in particular on scales below $\sim 20\, $\ts
arcmin which roughly demarcates the linear and nonlinear regimes. 
However it has been shown that the quasi-linear theory yields a
surprisingly accurate estimate of the skewness of the density field,
even in the regime 
($\abs{\delta}\sim 1$) where one could not expect the
perturbation series to yield reliable results (e.g. Baugh, Gazta\~naga \&
Efstathiou 1995, Colombi, Bouchet \&
Hernquist 1996, Gazta\~naga \& Bernardeau 1997). Therefore, we expect that
the results for $S(\theta)$ as calculated here will provide a good
approximation to the true skewness, even on scales down to a few
arcminutes. On very small scales with density contrasts much larger
than 1, the skewness of the density measured in N-body simulations
is indeed larger than the perturbation theory value, but by no more
than a factor of two (Baugh et al. 1995, Colombi et al. 1996).  

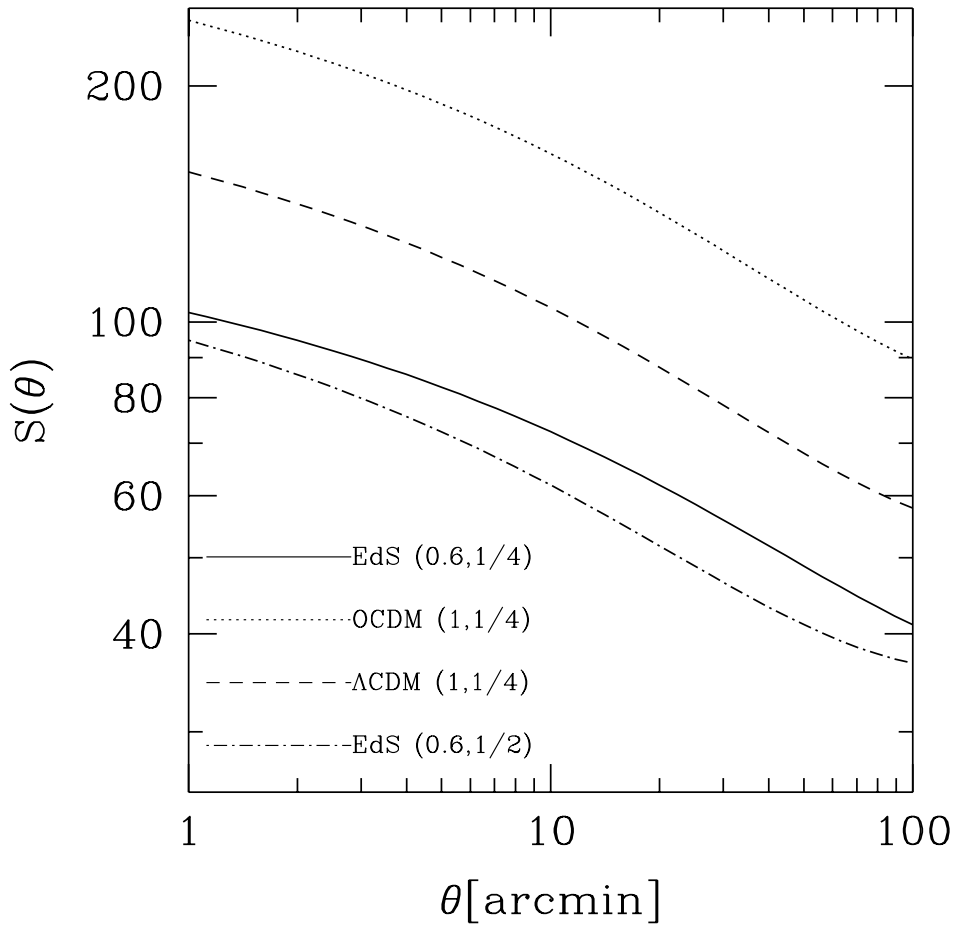
\xfigure{5}{The skewness, as defined in (4.12), as a function of
filter scale $\theta$, for four cosmological models. Since the
skewness as calculated from quasi-linear theory is independent of the
normalization of the power spectrum, the two EdS models with the same
$\Gamma$ yield the same $S(\theta)$ curves. The sources were
distributed according to (2.18), with $z_0=1$ and $\beta=1.5$, and the
filter with $\ell=1$ was used}{fig5.tps}{13}

In Fig.\ts 5 we have plotted the skewness as a function of angular
scale, for various cosmological models. Going from an EdS model to an
open model, the skewness increases by more than a factor of two, with
the spatially flat $\Lambda$ model taking intermediate values. The two EdS
models with different power spectra have only slightly different
skewness. This result is in full agreement with BvWM, who found that
the skewness is a fairly sensitive probe for the cosmological model,
to a large degree independent of the exact shape of the power
spectrum. The skewness on an angular scale of $5'$ is plotted in
Fig.\ts 6, for cosmological models with $\Omega_{\rm v}=0$, and flat
cosmologies ($\Omega_{\rm v}=1-\Omega_{\rm d}$), and two different
shape parameters $\Gamma$ of the power spectrum, as a function of
$\Omega_{\rm d}$. The two panels are for different source redshift
distributions. Again we see that the variation of $S$ with the
cosmological parameters is much stronger than with the shape of the
power spectrum. 
And by definition $S$ is independent of the normalization $\sigma_8$ 
of the power spectrum, at least in perturbation theory, where it is
the ratio of two terms, each of $\O\rund{\eckk{\delta^{(1)}}^4}$. 
We therefore conclude that the skewness is a powerful
discriminator between different cosmological models. 

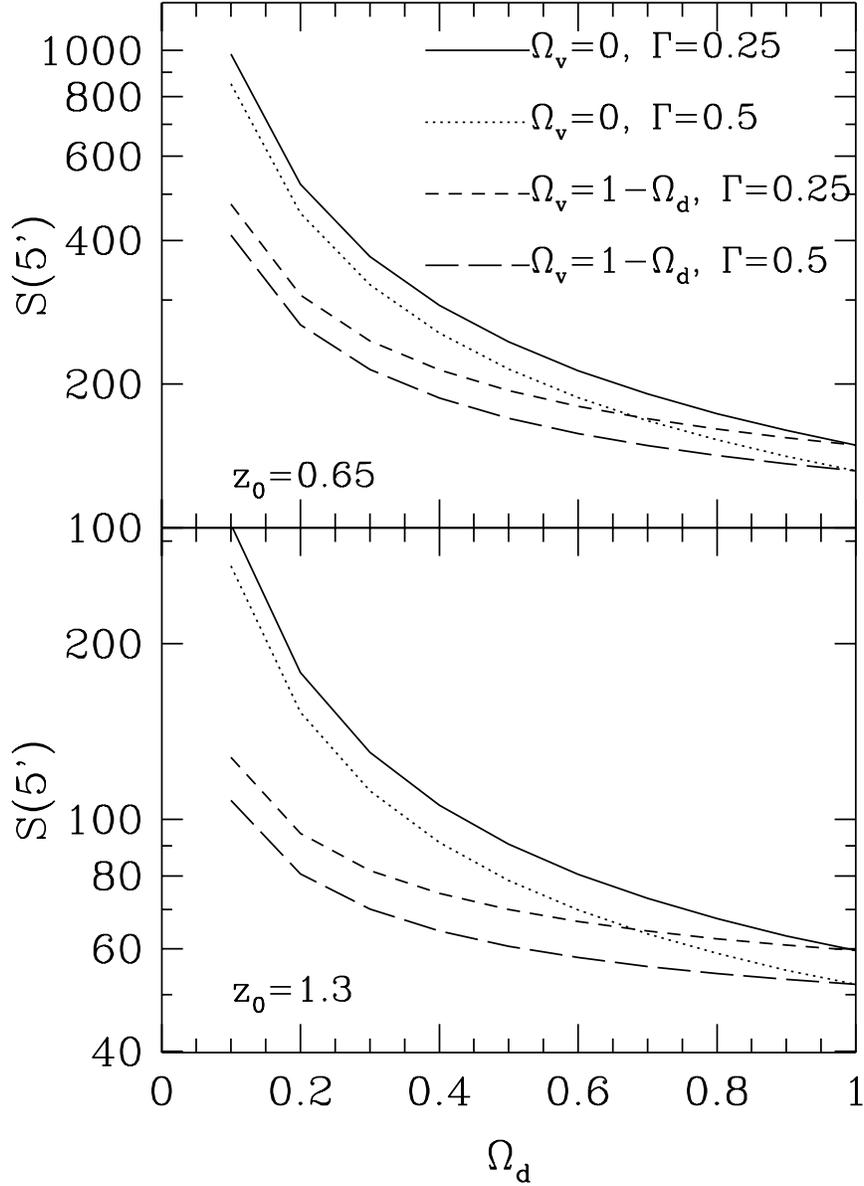
\xfigure{6}{The skewness, as defined in (4.12), as a function of
$\Omega_{\rm d}$, for vanishing cosmological constant $\Omega_{\rm
v}=0$, and for flat Universes, $\Omega_{\rm v}=1-\Omega_{\rm d}$. For
both cases, two values of the shape parameter $\Gamma$ were
considered. The redshift distribution of the sources was assumed to
follow (2.18), with $\beta=1.5$, and $z_0=0.65$ (upper panel) and
$z_0=1.3$ (lower panel), corresponding to a mean redshift of about one
and two, respectively. As can be seen, the dependence of $S$ on the
cosmological parameters is considerably stronger than the dependence
on the shape of the power spectrum}{fig6.tps}{13}

Since $\ave{\m^3}$ contains a factor $\Omega_{\rm d}^3$, and
$\ave{\m^2}$ has a factor $\Omega_{\rm d}^2$, one expects that to
leading order, $S\propto \Omega_{\rm d}^{-1}$. As can be seen from
Fig.\ts 6, $S$ follows this expectation rather closely for the models
with $\Omega_{\rm v}=0$. For flat cosmologies, however, these
$\Omega_{\rm d}$ factors are no longer the dominant terms in the
dependence of $S$ on $\Omega_{\rm d}$. The dependence on $\Omega_{\rm
d}$ is weakened due to the distance factors in the integrands for 
$\ave{\m^3}$ and $\ave{\m^2}$. 

Up to now, we have considered the skewness as it arises due to the
quasi-linear evolution of the density field which transforms an
initially Gaussian field into a non-Gaussian one. The linear density
field would not cause any skewness in the frame of the approximations
used up to now.  Nevertheless, even in the case of Gaussian density
fluctuations, the observable skewness would not be identically zero,
owing to the following three effects: (i) Light rays do not propagate
along `straight' lines, but are deflected. Therefore, the separation
vector $\vc x(\vc\theta,w)$ deviates from $f_K(w)\vc\theta$. This
deviation from the `Born approximation' leads to a non-vanishing
skewness. (ii) In the transition from (2.9) to (2.10), in addition to
the Born approximation the matrix $\A$ on the right-hand-side of (2.9)
was approximated by the unit matrix; in that way, the leading order term
(in $\Phi$) of the propagation matrix was obtained. The next
order term, which describes the coupling between lens planes, yields a
non-vanishing skewness. These two effects have been quantitatively
analyzed in BvWB for the skewness of the projected density field,
where it was shown that they are small compared to the expression
obtained from quasi-linear density evolution. (iii) The fact that the
image ellipticity is not an unbiased estimator for the shear $\gamma$,
but for the reduced shear $g=\gamma/(1-\kappa)$, implies that a
Gaussian density field can produce a finite skewness to the extent
that the weak lensing condition $\kappa\ll 1$ is no longer valid, and
that we have to consider the term $\gamma^{(1)}\kappa^{(1)}$. We shall
consider all three effects in the Appendix. All three effects yield
contributions to $\ave{\m^3}$ which are proportional to the square of
the power spectrum $P_0(k)$, as is the case for the leading term
considered here. Therefore, there is no a priori reason to expect that
these `correction' terms are much smaller than (4.11). As it turns out,
however, these correction terms amount to $\sim 5$\% of the leading
term (4.11), and may therefore be safely neglected presently.
%
%
%

A final point concerns the redshift distribution of the sources.
Fig.\ts 6 shows that the source redshift dependence of $S$ is rather
large, as pointed out by BvWM. This is a potential problem for two
reasons: (1) First, it means that in order to compare an observed
skewness with theoretical calculations, we have to know the redshift
distribution fairly precisely. Since the likely sources are very
faint, a complete spectroscopic survey to the corresponding magnitude
limits is not available, but progress in the method of photometric
redshifts may yield sufficiently accurate estimates of the required
distribution. (2) The magnification bias changes the redshift
distribution of the observed galaxy iamges in a way which depends on
the local value of the magnification. The observed skewness is
sensitive to this.  It is not the aim of this paper to discuss these
points. A rough estimate of the second point shows that the
magnification bias is negligible, but a detailed and systematic
analysis of these effects remains to be done.

\sec{5 Practical estimates}
We shall now consider some practical estimators for the
$\m$-statistics, and their dispersion. The results of this section
yield an estimate of the observations necessary to measure $\ave{\m^2}$
with given precision.

\subs{5.1 Dispersion for a single field}
Consider first a single circular field of angular radius $\theta$, in which
$N$ galaxies are observed at positions $\vc\theta_i$ with complex
ellipticity $\eps_i$. In the case of weak lensing, $\kappa\ll 1$, the
transformation from the intrinsic ellipticity $\eps^\s$ to the
observed one is simply $\eps=\gamma+\eps^\s$. Since the intrinsic
orientation of galaxies is random, $\eps$ is an unbiased estimator of
the local shear. Defining in analogy to (3.9) the tangential component
of the ellipticity of a galaxy at
$\vc\theta_i=(\theta_i\cos\vp_i,\theta_i\sin\vp_i)$ by
$$
\eps_{{\rm t}i}=-\Re \rund{\eps_i\,{\rm e}^{-2{\rm i}\vp_i}}\; ,
\eqno (5.1)
$$
then $\eps_{{\rm t}i}$ is an unbiased estimator of $\gamma_{\rm
t}(\vc\theta_i)\equiv \gamma_{{\rm t}i}$. In terms of the observable
image ellipticities, we define the estimator
$$
M:={(\pi \theta^2)^2\over N(N-1)}\sum_{i,j\ne i}^N Q_i Q_j \eps_{{\rm t}i}
\eps_{{\rm t}j} \;,
\eqno (5.2)
$$ 
where $Q_i\equiv Q(\vc\theta_i)$, which in turn is defined in (3.8), 
and the sum is taken only over terms
with $i\ne j$. The expectation value of $M$ is obtained by three
averaging processes: averaging over the intrinsic ellipticity
distribution, averaging over the galaxy positions, and the ensemble
average. We denote the first of these processes by the operator ${\tt
A}$, and the second by ${\tt P}$; these two operators commute. Then,
$$
{\tt A}(\eps_{{\rm t}i} \eps_{{\rm t}j})
={\tt A}\rund{\eckk{\gamma_{{\rm t}i}+ \eps_{{\rm t}i}^\s}\,
\eckk{\gamma_{{\rm t}j}+ \eps_{{\rm t}j}^\s}}
=\gamma_{{\rm t}i}\gamma_{{\rm t}j}+{\sigma_\eps^2\over
2}\delta_{ij}\;,
\eqno (5.3)
$$
since ${\tt A}(\eps_{{\rm t}i}^\s)=0$, owing to the random intrinsic
orientation; here, $\sigma_\eps$ is the dispersion of the intrinsic
ellipticity distribution. The term with $\sigma_\eps^2$ is divided 
by $2$ as we want the dispersion of only one component of the complex 
ellipticity.   In (5.3) we also used the fact that
the intrinsic ellipticity is uncorrelated with the shear. Note that
the second term in (5.3) does not appear in the sum of (5.2).
The operator ${\tt P}$ which averages over galaxy positions is defined as
$$
{\tt P}(X)=\rund{\prod_{i=1}^N \int{\d^2\theta_i\over \pi \theta^2}} X\;,
\eqno (5.4)
$$
so that 
$$
{\tt P}\rund{\sum_{i,j\ne i}^N Q_i Q_j \gamma_{{\rm t}i}\gamma_{{\rm
t}j}} ={N(N-1)\over (\pi \theta^2)^2}
\int \d^2\theta_1\;Q(\theta_1)\int \d^2\theta_2\; Q(\theta_2)\,
\gamma_{\rm t}(\vc\theta_1)\,\gamma_{\rm t}(\vc\theta_2)\; ,
\eqno (5.5)
$$
because the sum yields $N(N-1)$ equal terms.
The expectation value ${\rm E}(M)$ of $M$ then becomes:
$$
{\rm E}(M)\equiv \ave{{\tt P}({\tt A}(M))}
=\int \d^2\theta_1\;Q(\theta_1)\int \d^2\theta_2\; Q(\theta_2)\,
\ave{\gamma_{\rm t}(\vc\theta_1)\,\gamma_{\rm t}(\vc\theta_2)}
=\ave{\m^2}\; .
\eqno (5.6)
$$
Hence, $M$ is an unbiased estimator for $\ave{\m^2}$. 

Next, we consider the dispersion of this estimator,
$$
\sigma^2(M)={\rm E}(M^2)-\eckk{{\rm E}(M)}^2\; ,
\eqno (5.7)
$$
with
$$
M^2={(\pi \theta^2)^4\over N^2(N-1)^2}\sum_{i,j\ne i}^N Q_i Q_j \eps_{{\rm t}i}
\eps_{{\rm t}j} \sum_{k,l\ne k}^N Q_k Q_l \eps_{{\rm t}k}
\eps_{{\rm t}l} 
\;.
\eqno (5.8)
$$ 
Starting with performing the average over the intrinsic ellipticity
distribution, we find
$$\eqalign{
{\tt A}&(\eps_{{\rm t}i}\eps_{{\rm t}j}\eps_{{\rm t}k}\eps_{{\rm t}l} )
=\gamma_{{\rm t}i}\gamma_{{\rm t}j}\gamma_{{\rm t}k}\gamma_{{\rm t}l} 
+{\tt A}\rund{\eps_{{\rm t}i}^\s\eps_{{\rm t}j}^\s\eps_{{\rm
t}k}^\s\eps_{{\rm t}l}^\s }\cr
&+{\sigma_\eps^2\over 2}
\rund{\gamma_{{\rm t}i}\gamma_{{\rm t}j}\delta_{kl}
+\gamma_{{\rm t}i}\gamma_{{\rm t}k}\delta_{jl}
+\gamma_{{\rm t}i}\gamma_{{\rm t}l}\delta_{jk}
+\gamma_{{\rm t}j}\gamma_{{\rm t}k}\delta_{il}
+\gamma_{{\rm t}j}\gamma_{{\rm t}l}\delta_{ik}
+\gamma_{{\rm t}k}\gamma_{{\rm t}l}\delta_{ij} } \;. \cr }
\eqno (5.9)
$$
Owing to the restrictions in the sums of (5.8), the first and last term
in the second parenthesis do not contribute. 

To evaluate the position-average
over the first term in (5.9), we have to consider three different
cases: (a) all four indices $i, j, k, l$ are different; (b) one of the
second pair of indices $k,l$ is equal to one of the first pair $i,j$;
(c) both indices in the second pair are equal to those in the
first. These cases occur in the sum of (5.8) in $N(N-1)(N-2)(N-3)$,
$4N(N-1)(N-2)$, and $2N(N-1)$ terms, respectively. Therefore,
$$\eqalign{
{\tt P}&\rund{\sum_{i,j\ne i}^N\sum_{k,l\ne k}^N Q_i Q_j Q_k Q_l
\gamma_{{\rm t}i}\gamma_{{\rm t}j}\gamma_{{\rm t}k}\gamma_{{\rm t}l}}
\cr
&={N(N-1)(N-2)(N-3)\over (\pi \theta^2)^4}\,\m^4
+{4N(N-1)(N-2)\over  (\pi \theta^2)^3}\, \m^2
\int\d^2\vt\; Q^2(\vt)\,\gamma_{\rm t}^2(\vc\vt) \cr
&+{2N(N-1)\over (\pi \theta^2)^2} \rund{\int\d^2\vt\; 
Q^2(\vt)\,\gamma_{\rm t}^2(\vc\vt)}^2 \cr }
\eqno (5.10)
$$
For the second term in (5.9), we note that since $i\ne j$, $k\ne l$, 
${\tt A}\rund{\eps_{{\rm t}i}^\s\eps_{{\rm t}j}^\s\eps_{{\rm
t}k}^\s\eps_{{\rm t}l}^\s }$ contributes to the sum in (5.8) only for
$k=i$ and $l=j$, or $k=j$ and $l=i$. There are $2N(N-1)$ such
terms. Hence,
$$
{\tt P}\rund{\sum_{i,j\ne i}^N\sum_{k,l\ne k}^N Q_i Q_j Q_k Q_l
{\tt A}\rund{\eps_{{\rm t}i}^\s\eps_{{\rm t}j}^\s\eps_{{\rm
t}k}^\s\eps_{{\rm t}l}^\s }}
={2 N (N-1)\over (\pi \theta^2)^4}\rund{\sigma_\eps^2\over 2}^2
G^2\;,
\eqno (5.11)
$$
where we have defined
$$
G=\pi \theta^2 \int\d^2\vt\;Q^2(\vt)\; .
\eqno (5.12)
$$
Finally, each of the four terms in the last term of (5.9) which
contribute to the sum in (5.8) yield the same result when averaged
over position. Considering one of these, 
$$\eqalign{
{\tt P}\rund{\sum_{i,j\ne i}^N\sum_{k,l\ne k}^N Q_i Q_j Q_k Q_l
\gamma_{{\rm t}i}\gamma_{{\rm t}k}\delta_{jl}}
&={\tt P}\rund{\sum_{j,i\ne j,k\ne j}^N Q_i Q_j^2 Q_k 
\gamma_{{\rm t}i}\gamma_{{\rm t}k}} \cr
={N(N-1)(N-2)\over (\pi \theta^2)^4}\,\m^2 \, G
&+{N(N-1)\over (\pi \theta^2)^3} G \int\d^2\vt\; 
Q^2(\vt)\,\gamma_{\rm t}^2(\vc\vt) \;,\cr  }
\eqno (5.13)
$$
where the first term comes from terms with $i\ne k$, and the second
are those from $i=k$. 
Collecting terms, and taking the ensemble average yields for the
dispersion
$$\eqalign{
\sigma^2(M)&={(N-2)(N-3)\over N (N-1)}\ave{\m^4}
+{4 (N-2)\over N(N-1)}
\ave{\m^2 M_{\rm s}^2}\cr
&+{2\over N(N-1)}
\ave{M_{\rm s}^4} 
+{2(N-2)\over N(N-1)} \sigma_\eps^2\,G\, \ave{\m^2} \cr
&+{2\over N(N-1)}\sigma_\eps^2\, G \,\ave{M_{\rm s}^2}
+{1\over 2N(N-1)}\,\sigma_\eps^4\,G^2 -\ave{\m^2}^2\;, \cr }
\eqno (5.14)
$$
with
$$
M_{\rm s}^2:=\pi \theta^2 \int\d^2\vt\; 
Q^2(\vt)\,\gamma_{\rm t}^2(\vc\vt)\; .
\eqno (5.15)
$$
Note that the dispersion contains three different sources of noise:
the contribution due to the finite width of the intrinsic ellipticity
distribution, the `cosmic variance', and the noise due to the finite
number of randomly located 
galaxy images. The latter effect is contained in the
terms which include $M_{\rm s}$. However, this
source of noise is never dominant: since the magnitude of $M_{\rm s}^2$
will be comparable with that of $\m^2$, the second term in (5.14) is
smaller by a factor $1/N$ than the first, and smaller by a factor $2G
M_{\rm s}^2/\sigma_\eps^2$ than the fourth term. Similar estimates are
valid for the other two terms containing $M_{\rm s}^2$. Thus, by dropping
terms containing $M_{\rm s}$ and taking $N\gg 1$, we find
$$
\sigma^2(M)\approx \mu_4 \ave{\m^2}^2 +\rund{{\sigma_\eps^2 G\over
\sqrt{2} N} + \sqrt{2}\ave{\m^2}}^2\; ,
\eqno (5.16)
$$
where $\mu_4$ is the kurtosis of $\m$, $\mu_4=\ave{\m^4}/\ave{\m^2}^2-3$,
which vanishes if $\m$ is distributed like a Gaussian.
We note that $G$ is a factor of order unity; if we choose the weight
function (3.12 \& 13), we obtain
$$
G={(1+\ell)(2+\ell)^2 \over (1+2\ell)(3+2\ell)}\;,
\eqno (5.17)
$$
which yields $G=6/5$, $48/35$, and $100/63$ for $\ell=1$, 2, 3. In
order to see whether the cosmic variance or the intrinsic ellipticity
distribution dominates the noise, we consider the ratio
$$
\rho={2\ave{\m^2} N\over \sigma_\eps^2 G}
={1500 \pi\over G}\ave{\m^2(\theta)}\rund{\theta\over 1\,{\rm arcmin}}^2
\rund{\sigma_\eps\over 0.2}^{-2}\rund{n\over 30{\rm arcmin}^{-2}}\;,
\eqno (5.18)
$$
where $n=N/(\pi \theta^2)$ is the mean density of galaxy images. If
$\rho\ll 1$, the intrinsic ellipticity distribution contributes mostly
to the noise, whereas in the other case, the cosmic variance is the
dominating factor. In this discussion the kurtosis term in (5.16) has
been ignored. 

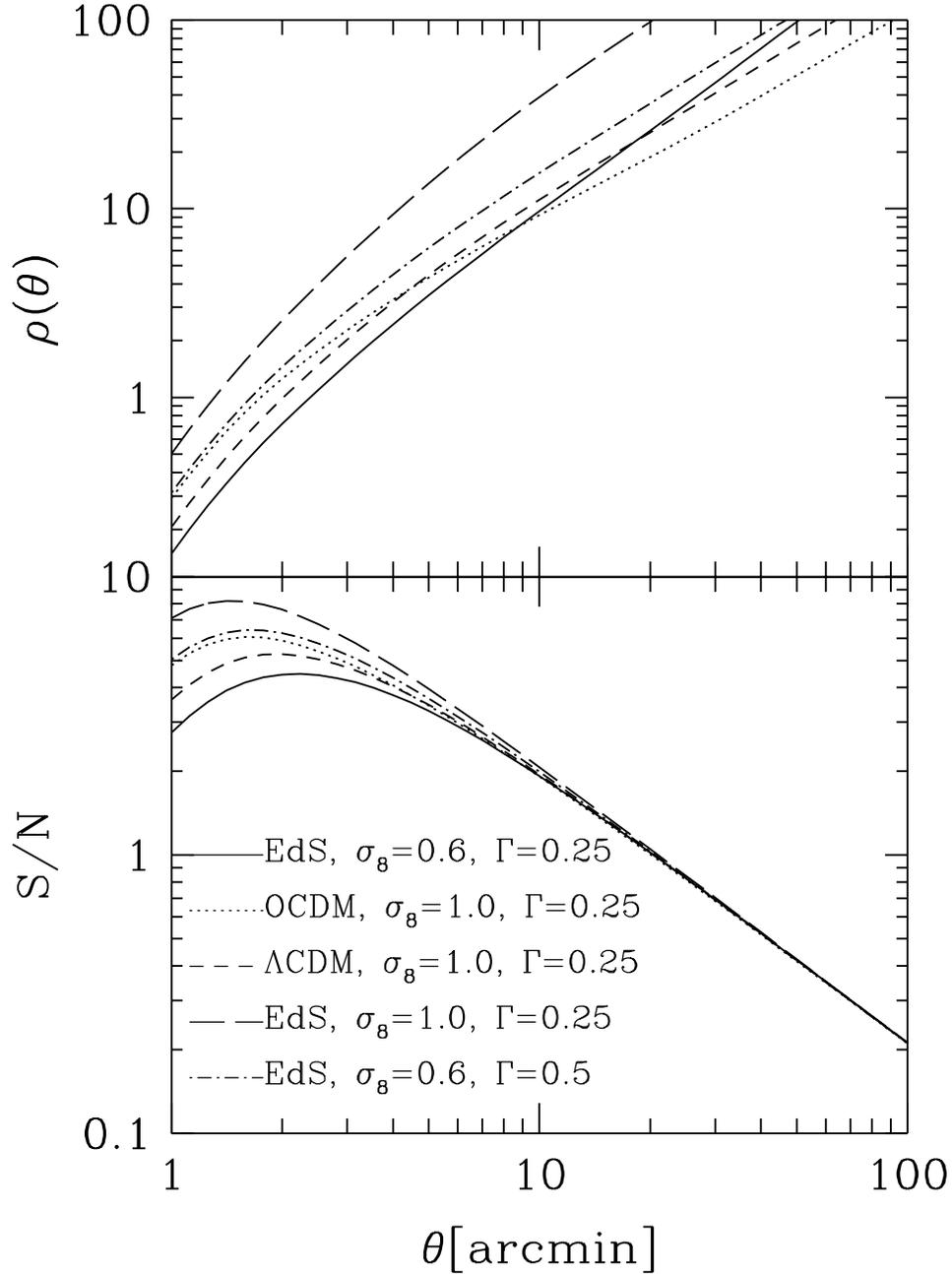
\xfigure{7}{The upper panel displays the quantity $\rho$ -- see (5.18)
-- as a function
of angle $\theta$, which describes the relative importance of the
cosmic variance relative to the 
intrinsic galaxy ellipticity dispersion, for the same cosmological
models as shown in Fig.\ts 1. In this figure, we have set
$\sigma_\eps=0.2$ and $n=30{\rm arcmin}^{-2}$; the dependence on these
two quantities can be seen from (5.18). Also, we have used the filter with
$\ell =1$. For all cosmological models considered here, the cosmic
shear dominates for angular scales above a few arcminutes. In the
lower panel, we plot the signal-to-noise ratio (5.22) for the
dispersion on angular scale $\theta$, for the case that one has a
square field of length $L=1$\ts degree available, and that
this field is densely 
covered with circles of radius $\theta$, so that $N_{\rm
f}=(L/2\theta)^2$. We have assumed zero kurtosis for this plot.
The signal-to-noise ratio at fixed $\theta$
is proportional to $L$. Parameters were chosen as in the
upper panel. For both panels, the redshift distribution was chosen
according to (2.18), with $z_0=1$ and $\beta=1.5$, and the fully
non-linear power spectrum was used}{fig7.tps}{14} 

The coefficient $\rho$ is plotted in the upper panel of Fig.\ts 7, as
a function of $\theta$, for the same cosmological models as considered
before. One sees that for filter scales exceeding a few arcminutes,
the intrinsic ellipticity distribution of galaxies is no longer the
dominant source of noise, but the `cosmic variance' will start to
dominate.  

\subs{5.2 Dispersion for an ensemble of fields}
Obviously, from observing a single field with radius $\theta$, no reliable
estimate for $\ave{\m^2(\theta)}$ can be obtained. We next consider a sample
of $N_{\rm f}$ fields, and assume that they are spatially sufficiently
well separated so that the shear in one field is statistically
independent of that in the others. In that case, an unbiased estimator
of $\ave{\m^2}$ is provided by
$$
\M=\rund{\sum_{n=1}^{N_{\rm f}} a_n}^{-1}
\sum_{n=1}^{N_{\rm f}} a_n\,M_n\;,
\eqno (5.19)
$$
where $M_n$ is the estimator (5.2) in a single field, and the $a_n$
are weight factors. It is easy to show that the dispersion of $\M$ is
minimized if $a_n\propto \sigma^{-2}(M_n)$, in which case it becomes
$$
\sigma^2(\M)=\rund{\sum_{n=1}^{N_{\rm f}} \sigma^{-2}(M_n)}^{-2}
\sum_{n=1}^{N_{\rm f}}  \sigma^{-2}(M_n)\; .
\eqno (5.20)
$$
If all $N_{\rm f}$ fields contain the same number of galaxies, then,
as expected,
$$
\sigma(\M)={\sigma(M)\over \sqrt{N_{\rm f}}}\; .
\eqno (5.21)
$$
Combining (5.16) with (5.18) and (5.21), we find for the
signal-to-noise ratio for a measurement of ${\ave{\m^2}}$:
$$
{{\rm S}\over {\rm N}}:= {{\ave{m^2}}\over \sigma(\M)}
={\sqrt{N_{\rm f}}\over \sqrt{\mu_4+2\rund{1+{1\over \rho}}^2}}\;.
\eqno (5.22)
$$

The question of how far two fields have to be separated before they
can be considered statistically independent can be investigated by
considering the correlation between the values of $\m$ in two fields of
angular radius $\theta_1$ and $\theta_2$, 
separated by $\Delta\theta$. We consider the correlation
coefficient
$$
r_{\rm corr}(\theta_1,\theta_2;\Delta\theta):={\ave{\m(\theta_1)
\m(\theta_2)}(\Delta\theta) \over
\sqrt{\ave{\m^2(\theta_1)}\ave{\m^2(\theta_2)}}}\;,
\eqno (5.23)
$$
which can be easily calculated with (3.5) and (3.2) to yield
$$
r_{\rm corr}(\theta_1,\theta_2;\Delta\theta)={2\pi\over
\sqrt{\ave{\m^2(\theta_1)}\ave{\m^2(\theta_2)}}} 
\int_0^\infty \d
s\;s\,P_\kappa(s)\,J_0(s\,\Delta\theta)\,I_\ell(s \theta_1)\,I_\ell(s
\theta_2)\; .
\eqno (5.24)
$$
This can be compared with the correlation of the mean shear
$\bar\gamma(\theta)$ as defined in (3.16), for which the corresponding
correlation coefficient is the same as in (5.24) with $I_\ell(s R)$
replaced by $I_{\rm TH}(s \theta)$. In Fig.\ts 8 we have plotted
$r_{\rm corr}$ for various values of $\theta_1$, $\theta_2$ and
$\Delta\theta$, both for the $\m$-statistics and for $\bar\gamma$. From
the figure is can be easily seen that the $\m$-statistics decorrelates
very quickly. For example, considering two circles of equal radius,
such that they just touch (so that
$\theta_1=\theta_2=\Delta\theta/2$), we infer from the upper left
panel in Fig.\ts 8 that the correlation between the $\m$ measurements in
these two apertures is less than 1\%! Therefore, if we had a large
image, we can place apertures densely on that image and
consider the $\m$-values obtained from each aperture as
independent. Also, for different values of $\theta_1$ and $\theta_2$,
the decorrelation is very quick. This property can then be employed for
obtaining $\ave{\m^2(\theta)}$ from a big image for various angular
scales. 

Thus, for an image with sidelength $L$, we can place
$N_{\rm f}=(L/2\theta)^2$ nearly independent apertures on this
field. Using this estimate, we have plotted S/N, as defined in (5.22),
in the lower panel of Fig.\ts 7, assuming $L=1$\ts degree.  There we
can see that  the S/N ratio is larger than one for angular scales
below $20'$. For different values of $L$, one uses the scaling ${\rm
S/N}\propto L$. The results of Fig.\ts 7 can be compared with the
Fig.\ts 10 of JS which gives the signal-to-noise of the rms shear
with a top-hat window. 

In contrast to $\m$, $\bar\gamma$ decorrelates very slowly, as
indicated by the light curves in Fig.\ts 8. This can be traced
directly to the shape of the top-hat filter $I_{\rm TH}$, displayed in
Fig.\ts 2, which picks up long-scale power of the density
field. Therefore, if one wants to measure cosmic shear using
$\bar\gamma$, one has to use data fields which are well separated on
the sky.

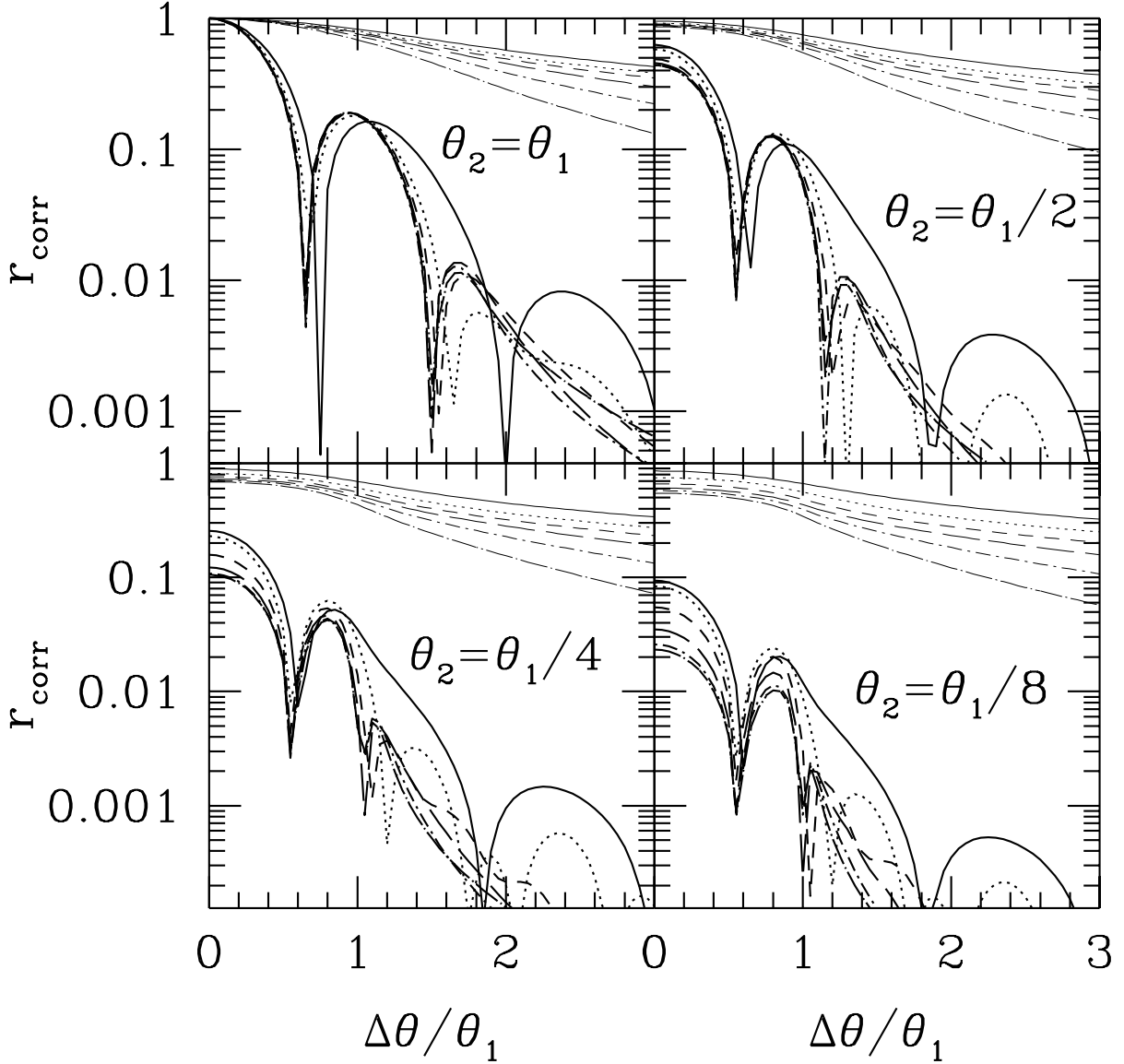
\xfigure{8}{The correlation coefficient $r_{\rm
corr}(\theta_1,\theta_2;\Delta\theta)$ is plotted as a function of
$\Delta\theta/\theta_1$ (heavy curves). Here, $\theta_1$ is the larger
of the two aperture sizes; the other one is 1, 1/2, 1/4, and 1/8 of
the former one in the four different panels. Six curves are drawn in
each panel, corresponding to $\theta_1/{\rm arcmin}=1$ (solid curves), 2 (dotted
curves), 4 (short-dashed curves), 8 (long-dashed curves), 16
(short-dashed-dotted curves), and 32 (long-dashed-dotted curves). The
light curves in each panel is the corresponding correlation
coefficient for the mean shear inside circles,
$\bar\gamma(\theta)$. For this figure, an $\Omega_{\rm d}=1$,
$\Omega_{\rm v}=0$ universe has been assumed, with $\sigma_8=0.6$ and
$\Gamma=0.25$, and the sources were assumed to follow the redshift
distribution (2.18) with $z_0=1$ and $\beta=1.5$}{fig8.tps}{16} 

\subs{5.3 Alternative estimators}
An apparently unrelated estimator for $\ave{\m^2}$ can be obtained from
the following consideration: define
$$
\hat\gamma(\theta):=\int\d^2\vt\;U(\vt)\,\gamma(\vc\vt)\;,
\eqno (5.25)
$$
where the integral extend over a circle with radius $\theta$. This
definition is analogous to the one in (3.16), except that a weight
function is added. Then
$$
\ave{\hat\gamma \hat\gamma^*}
=\int\d^2\theta'\;U(\theta')\int\d^2\vt\;U(\vt)
\ave{\gamma(\vc\theta')\gamma^*(\vc\vt)}\;.
\eqno (5.26)
$$
Noting that the two-point correlation function of the shear is the
same as the two-point correlation function of $\kappa$ (see, e.g.,
Blandford et al.\ts 1991), we see that this expression agrees with
(3.10), so that
$$
\ave{\m^2(\theta)}=\ave{\hat\gamma(\theta) \hat\gamma^*(\theta)}\;.
\eqno (5.27)
$$
Hence, a practical estimator for $\ave{\m^2}$ is
$$
\hat M={(\pi \theta^2)^2\over N(N-1)}\sum_{i,j\ne i}^N
U(\theta_i)\,U(\theta_j)\,\rund{\eps_{1i}\eps_{1j}+\eps_{2i}\eps_{2j}}
\;.
\eqno (5.28)
$$
With calculation similar to those in Sect.\ts 5.1, one can show that
indeed $\hat M$ is an unbiased estimator for $\ave{\m^2}$; its
dispersion in the same approximation as in (5.16) is
$$
\sigma^2(\hat M)=\hat\mu_4 \ave{\m^2}^2
+\rund{{\sigma_\eps^2\,\hat G\over N}+\sqrt{2}\ave{\m^2}}^2\;,
\eqno (5.29)
$$
where $\hat\mu_4$ is the kurtosis of the shear $\hat\gamma$, and 
$$
\hat G=\pi \theta^2\int\d^2 \vt\; U^2(\vt) =G\;,
\eqno (5.30)
$$
where the final equality is valid if $U$ is chosen as in (3.12). Thus
we see that the dispersion of $\hat M$ is very similar to that of $M$,
except for a factor $\sqrt{2}$ in the term containing
$\sigma_\eps$. This difference is due to the fact that in $\hat M$,
two components of $\eps$ are used, each of which carries its
dispersion. Hence, for measuring $\ave{\m^2(\theta)}$, one can use the
estimators (5.2) and (5.28), where the former should yield a slightly
higher signal-to-noise ratio. A comparison between these two estimates
can be used to check the integrity of the data and the data analysis
procedure. 

Another check on the quality of the data can be obtained by noting
that 
$$
M_{\rm r}:=\int\d^2\vt\;Q(\abs{\vc\vt})\,\gamma_{\rm r}(\vc\vt) 
\eqno (5.31)
$$
should vanish identically, where $\gamma_{\rm r}$ is the radial
component of the shear, defined by taking the imaginary part in (3.9)
instead of the real part for the tangential component. The fact that
$M_{\rm r}=0$ can be shown easily by introducing the Fourier transform
of $\gamma_{\rm r}$ to obtain a result similar to (A5), with the
cosine replaced by a sine; the polar angle integral in (5.31) then
yields zero. The estimator
$$
\M_{\rm r}=\sum_{n=1}^{N_{\rm f}}\rund{ {(\pi \theta)^2\over
N(N-1)}\sum_{i,j\ne i}^N Q_i 
Q_j\,\eps_{{\rm r}i}\,\eps_{{\rm r}j}}
$$ 
should therefore yield a value within the dispersion, which is 
given by $\sigma(\M_{\rm r})=\sigma_\eps^2 G/(\sqrt{2 N_{\rm f}} N)$. 

\subs{5.4 Practical estimator for $\ave{\m^3}$}
An obvious estimator for $\ave{\m^3(\theta)}$ for a single field is
$$
M_3:={(\pi \theta^2)^3\over N^3}\sum_{i,j,k=1}^N Q_i Q_j Q_k
\eps_{{\rm t}i}\eps_{{\rm t}j}\eps_{{\rm t}k} \; .
\eqno (5.32)
$$
It is easy to show, using the methods used before, that ${\rm
E}(M_3)=\ave{\m^3}$. An estimator for an ensemble of fields can then be
defined immediately in analogy to (5.19).

\sec{6 Discussion and conclusions}
In this paper we have introduced the $\m$-statistics, or mass-aperture
statistics, as a new measure for cosmic shear. We have compared those
statistics with others suggested earlier, namely the shear
two-point correlation function, and the rms of the shear averaged over
circles. The main results can be summarized as follows:

(1) $\m$ is defined as a filtered version of the projected density
field $\kappa$, but can be calculated in terms of the tangential shear
inside a circular aperture. Hence, $\m$ has a well-defined physical
interpretation, and at the same time can be expressed in terms of the
observable shear. 

(2) The dispersion of $\m$, $\ave{\m^2(\theta)}$, can be expressed as a
convolution of the power spectrum of the projected density field with
a filter function which is strongly peaked. In contrast, the
corresponding filter function for the rms of the shear averaged over
circles is considerably broader and in particular picks up the long
wavelength range of the power spectrum. Therefore, the $\m$-statistics
has the nice property of being a well localized measure for
the projected density field. From measurements on different angular
scales, the power spectrum of the projected density field can be
constructed. A different method to obtain a local estimate of the
power spectrum of the projected density field has recently been
suggested by Kaiser (1996).

(3) In contrast to the mean shear inside a circle, $\m$ is a scalar
quantity. It is therefore possible to define the skewness of $\m$ which
is measurable. 

(4) Whereas $\ave{\m^2(\theta)}$ is smaller than
$\ave{\abs{\bar\gamma(\theta)}^2}$, at least on small scales, the
dispersion of the estimator for $\ave{\m^2}$ is smaller by a factor of
$\sqrt{2}$ than that for $\ave{\abs{\bar\gamma(\theta)}^2}$ on small
scales, due to the fact that in the former case, only one component of
the shear is needed for the estimate. On scales beyond a few
arcminutes, the dispersion of the estimates for both statistics is
dominated by the cosmic variance, so that the fractional accuracy of
the estimates is nearly $N_{\rm f}^{-1/2}$, where $N_{\rm f}$ is the
number of independent circular apertures.

(5) The values of $\m$ for neighboring circles decorrelates very
rapidly with increasing separation. In particular, we found that the
values of $\m$ calculated on circles which just touch are nearly
mutually independent, with a correlation coefficient below
$10^{-2}$. This implies that one can make use of wide-field images by
`punching' circles on these images and considering the $\m$ values
on these circles as mutually independent. In contrast to that, the
mean shear inside neighboring circles is very strongly correlated, and
these have to be separated by several degrees before they can be
considered independent. Therefore, the $\m$-statistics provides a very
valuable tool for measuring cosmic shear on scales of arcminutes from
wide-field images. A single one square degree image of sufficient
depth (e.g., corresponding to a few hours integration time on a
4-meter-class telescope) and high imaging quality can be used to
obtain $\ave{\m^2(\theta)}$ on scales below $\sim 5'$ with a signal-to-noise
larger than $\sim 3$. The signal-to-noise is proportional to the
square root of the solid angle covered by the available data.  

(6) The skewness of $\m$ is very sensitive to the cosmological model,
but rather independent of the shape of the power spectrum. In
particular, in the quasilinear regime, the skewness
is independent of the normalization of the power spectrum. Whereas the
latter property may not be strictly preserved when considering the
skewness in a fully non-linearly evolved density field, the experience
from comparing numerical simulations of the density field with
quasi-linear predictions has shown that at least the skewness of the
density field is rather well approximated by quasi-linear theory, even
for density contrasts of order unity (Baugh et al \ 1995,
Colombi et al.\ 1996, Gazta\~naga \& Bernardeau 1997). 
We therefore expect that the
independence of the skewness on the normalization will be roughly
preserved in the non-linear case. This makes
the skewness a very valuable probe for cosmological parameters.

(7) The difference between the predictions from linear evolution of
the cosmic density fluctuations, and the fully non-linear evolution of
the power spectrum, is much larger for the $m$-statistics than for the
mean shear within circles -- see Fig.\ts 3. This is related to the
fact that the filter function for $\ave{\m^2}$ is narrowly peaked,
whereas the corresponding filter for
$\ave{\abs{\bar\gamma(\theta)}^2}$ has a long tail towards long
wavelengths. Since the long wavelengths have not become non-linear
(see Fig.\ts 1), the non-linear effects for
$\ave{\abs{\bar\gamma(\theta)}^2}$ are necessarily weaker than for
$\ave{\m^2}$. 

The signal-to-noise estimates given in this paper suffer from the lack
of knowledge on the kurtosis of $\m$, which we have not attempted to
calculate, or more generally, the distribution function of $\m$.
Therefore, the estimate presented in Fig.\ts 7 may be
slightly optimistic. A more accurate estimate can be obtained from
numerical simulations, by taking high-resolution realizations of the
cosmic mass distribution and studying light propagation through such a
Universe. Such studies have been undertaken in various ways and with
various scientific goals in the past (e.g., Jaroszy\'nski et al.\ts
1990, Lee
\& Paczy\'nski 1990, 
Bartelmann \& Schneider 1991, Jaroszy\'nski 1991, 1992, Wambsganss et al.\
1995, 1997). In particular, Blandford et al. (1991) have compared
their analytical estimates for the rms shear inside circles with
results from numerical ray propagation simulations. Further work in
this direction will most useful. In particular, the
performance of practical estimators for the dispersions and skewness
can be evaluated, and the accuracy of the approximations used here
(i.e., the Born approximation and the neglect of lens-lens terms in
(2.10)) can be checked.

In summary, the $\m$-statistics appears to be an attractive measure
for cosmic shear. We expect that its applications can be
expanded beyond the range considered here. It is a promising method to
search for mass concentrations in the Universe, as pointed out in
Schneider (1996). A further application may include the correlation of
shear, as measured by $\m$, with the distribution of foreground
galaxies, measured through the same filter function $U$. We shall
exploit this route as a method for measuring the bias parameter, and
its dependence on scale, in a later publication.

We are grateful to many colleagues for discussions on the topic of
this paper; in particular, we want to thank M.\ts Bartelmann, F.\ts
Bernardeau, S.\ts Mao, Y.\ts Mellier, S.\ts Seitz and S.\ts White for
many helpful suggestions and discussions, and M.\ts Bartelmann and
S.\ts White for a careful reading of the manuscript.
This work was supported by the ``Sonderforschungsbereich 375-95 f\"ur
Astro--Teil\-chen\-phy\-sik" der Deutschen
For\-schungs\-ge\-mein\-schaft, by the Programme National de
Cosmologie of the Centre National de la Recherche Scientifique in
France, and by EU contract CHRX CT930120.

\sec{Appendix}
In this Appendix we calculate the contributions to the (observable)
skewness which would be present even in the case of a Gaussian density
field. The three effects which cause a non-zero value for $\ave{\m^3}$
also for Gaussian fields are: (1) Image ellipticities provide an unbiased
estimate of the reduced shear $g=\gamma/(1-\kappa)$, rather than the
shear $\gamma$ itself. (2) The separation vector $\vc x(\vc\theta,w)$
deviates from $f_K(w)\vc\theta$ -- see (2.7); the (`Born') approximation
leading from (2.9) to (2.10) neglects this effect. (3) Also, in the
same step, the matrix $\A$ on the right-hand-side of (2.9) was
approximated by the unit matrix; therefore, (2.10) does not contain
the coupling between deflectors at different redshifts. The two latter
terms have already been discussed, in somewhat different context, in
BvWM. We shall consider here these effects in turn. In this Appendix,
$\delta$ and $\Phi$ are meant 
to be the {\it linear} density field and its corresponding gravitational
potential, since we are interested in the contribution to the skewness
coming from the above-mentioned effects in the presence of a Gaussian
field. 

\subs{A1 Aperture mass in terms of $g$}
Let 
$$
g_{\rm t}(\vc\vt)={\gamma_{\rm t}(\vc\vt)\over 1-\kappa(\vc\vt)}
\eqno (A1)
$$
be the tangential component of the reduced shear. Then, from image
ellipticities, one obtains an unbiased estimate of
$$
M_g(\theta):=\int \d^2\vt\;Q(\abs{\vc\vt})\,g_{\rm t}(\vc\vt)
=\m(\theta)+\delta \m(\theta)\; ,
\eqno (A2)
$$
where
$$
\delta
\m(\theta)\approx \int\d^2\vt\;Q(\abs{\vc\vt})\,\kappa(\vc\vt)\,\gamma_{\rm
t}(\vc\vt)\; .
\eqno (A3)
$$
Hence, the observable skewness becomes 
$$
\ave{M_g^3(\theta)}\approx \ave{\m^3(\theta)}+3\ave{\m^2(\theta)\,\delta
\m(\theta)} \equiv \ave{\m^3(\theta)}+\delta\ave{\m^3(\theta)}_g\;.
\eqno (A4)
$$
Using the fact that the Fourier transform of the shear is
$\tilde\gamma(\vc s)=\tilde\kappa(\vc s) \, {\rm e}^{2{\rm i}\vp_s}$,
where $\vp_s$ is the polar angle of $\vc s$,
the tangential shear -- see (3.9) -- can be Fourier-decomposed as
$$
\gamma_{\rm t}(\vc\vt)=-\Re\rund{\int {\d^2 s\over (2\pi)^2}\,
{\rm e}^{2{\rm i}\vp_s}\,\tilde\kappa(\vc s)\,{\rm e}^{{\rm i}\vc
s\cdot\vc\vt} \,{\rm e}^{-2{\rm i}\vp}}
=-\int {\d^2 s\over (2\pi)^2}\,\cos 2(\vp-\vp_s)\,{\rm e}^{{\rm i}\vc
s\cdot\vc\vt}\,\tilde\kappa(\vc s)\; ,
\eqno (A5)
$$
where in the last step use was made of the fact that $\kappa(\vc\vt)$ is
real, and $\vp$ is the polar angle of $\vc\vt$. Then, from the
definition in (A4) and by introducing the expression (3.5) for
$\m(\theta)$, one obtains after replacing $\kappa(\vc\vt)$ by its Fourier
representation:
$$\eqalign{
\delta\ave{\m^3(\theta)}_g&=-3\int\d^2\vt_1\;U(\abs{\vc\vt_1})
\int {\d^2 s_1\over (2\pi)^2}\,{\rm e}^{{\rm i}\vc s_1\cdot\vc\vt_1}
\int\d^2\vt_2\;U(\abs{\vc\vt_2})
\int {\d^2 s_2\over (2\pi)^2}\,{\rm e}^{{\rm i}\vc s_2\cdot\vc\vt_2}\cr
\times  \int\d^2\vt & \;Q(\abs{\vc\vt})
\int {\d^2 s'\over (2\pi)^2}\,{\rm e}^{{\rm i}\vc s'\cdot\vc\vt}
\int {\d^2 s\over (2\pi)^2}\,\cos 2(\vp-\vp_s)\,{\rm e}^{{\rm i}\vc
s\cdot\vc\vt}
\ave{\tilde\kappa(\vc s_1)\tilde\kappa(\vc s_2)
\tilde\kappa(\vc s)\tilde\kappa(\vc s')} \; . \cr }
\eqno (A6)
$$
We calculate the correction $\delta\ave{\m^3}_g$ to leading order and
thus use the linear evolution of the cosmic density fluctuations. In
this approximation the density field remains Gaussian, and since
$\kappa$ is a linear functional of the density fluctuations, 
$\kappa$ is Gaussian. Then the correlator in (A6) becomes
$$\eqalign{
\ave{\tilde\kappa(\vc s_1)\tilde\kappa(\vc s_2)
\tilde\kappa(\vc s)\tilde\kappa(\vc s')} 
&=(2\pi)^4 P_\kappa(s_1)\,P_\kappa(s')\,\delta_{\rm D}(\vc s_1+\vc
s_2)\,\delta_{\rm D}(\vc s'+\vc s) \cr
&+ \hbox{2 terms obtained from permutation}\; . \cr }
\eqno (A7)
$$
Inserting (A7) into (A6), one notices that the first term yields zero,
whereas the other two terms yield equal contributions. After
carrying out the $\vc s_1$ and $\vc s_2$ integrations, one finds
$$\eqalign{
\delta\ave{\m^3(\theta)}_g&=-6\int\d^2\vt_1\;U(\abs{\vc\vt_1})
\int\d^2\vt_2\;U(\abs{\vc\vt_2})  \int\d^2\vt  \;Q(\abs{\vc\vt})\cr
\times\int {\d^2 s\over (2\pi)^2} & \,\cos 2(\vp-\vp_s)\,{\rm e}^{{\rm i}\vc
s\cdot(\vc\vt-\vc\vt_1)} P_\kappa(s)
\int {\d^2 s'\over (2\pi)^2}\,{\rm e}^{{\rm i}\vc
s'\cdot(\vc\vt-\vc\vt_2)} P_\kappa(s')\; . \cr }
\eqno (A8)
$$
Next, we perform the polar angle integration of $\vc\vt_1$ and
$\vc\vt_2$, and use the definition of the filter $I_\ell(\eta)$ to obtain
$$\eqalign{
\delta\ave{\m^3(\theta)}_g&=-6 \int\d^2\vt  \;Q(\abs{\vc\vt})
\int {\d^2 s\over (2\pi)^2}\,\cos 2(\vp-\vp_s)\,{\rm e}^{{\rm i}\vc
s\cdot\vc\vt} P_\kappa(s)\,I_\ell(s\theta) \cr
&\times \int\d^2 s' {\rm e}^{{\rm i}\vc
s'\cdot\vc\vt} P_\kappa(s')\,I_\ell(s'\theta) \;. \cr }
\eqno (A9)
$$
Finally, the three integrations over the polar angles can be carried
out, leaving the final result
$$\eqalign{
\delta\ave{\m^3(\theta)}_g=12\pi&\int\d s\;s\,P_\kappa(s)\,I_\ell(s\theta)
\int \d s'\; s'\,P_\kappa(s')\,I_\ell(s'\theta)\cr
&\times \int_0^\theta \d\vt\;\vt\,Q(\vt)\,{\rm J}_0(\vt s')\,{\rm J}_2(\vt
s)\; . \cr }
\eqno (A10)
$$
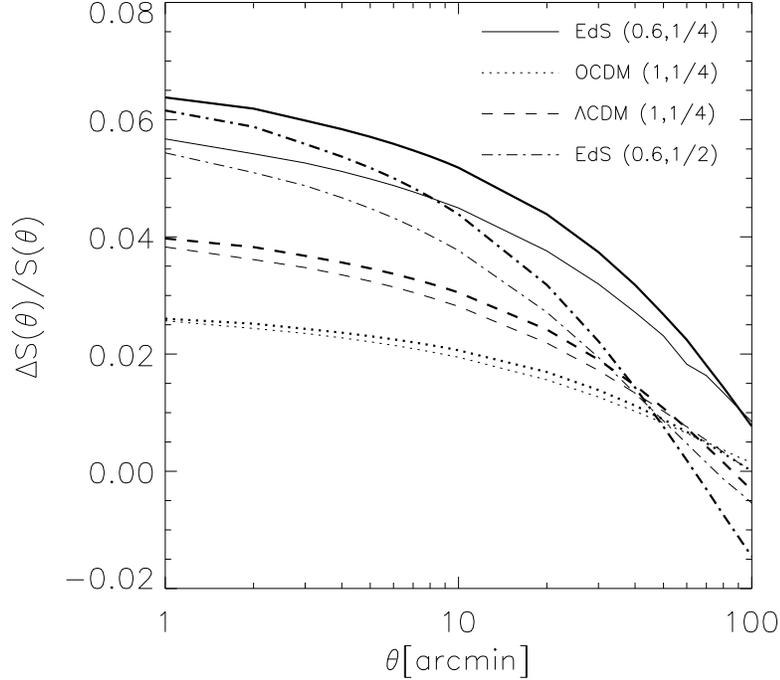
\xfigure{9}{The fractional change of the skewness which occurs by
considering $\ave{\m^3}$ instead of $\ave{M_g^3}$, where the latter one
is the observable quantity, for four different cosmological models
(thick curves). For comparison, the thin curves show the corrections
due to the Born approximation and the neglect of lens-lens coupling
terms in (2.10). In this figure, the sources were assumed to follow
the redshift distribution (2.18), with $z_0=1$ and $\beta=1.5$}
{fig9.tps}{13}

In Fig.\ts 9 we have plotted the fractional contribution
$\delta\ave{\m^3(\theta)}_g/ \ave{\m^3(\theta)}$ as a function of angular
scale $\theta$, for four different cosmological models. As can be
seen, the difference between $\ave{M_g^3}$ and $\ave{\m^3}$ is very small,
lower than 6\% for all cases considered, where the largest deviations
occur in the EdS models. 

\subs{A2 Lens-lens coupling, and dropping the Born approximation}
The expression (2.10) for the Jacobi matrix $\A$ is valid up to first
order in the gravitational potential $\Phi$. The two effects
considered here are obtained by expanding (2.9) up to second order in
$\Phi$. 

We write (2.7) as $\vc x(\vc\theta,w)=\vc x^{(0)}(\vc\theta,w)+\vc
x^{(1)}(\vc\theta,w)+\O(\Phi^2)$, with $\vc x^{(0)}(\vc\theta,w) =
f_K(w)\vc\theta$ and $\vc x^{(1)}(\vc\theta,w)$ given by the second
term in (2.7), now with $\vc x$ in the integrand replaced by $\vc
x^{(0)}$. Similarily, we write $\A(\vc\theta,w)=\A^{(0)}(\vc\theta,w)
+\A^{(1)}(\vc\theta,w)+\A^{(2)}(\vc\theta,w)+\O(\Phi^3)$, with
$\A_{ij}^{(0)}(\vc\theta,w)=\delta_{ij}$, and $\A^{(1)}(\vc\theta,w)$
given by the second term of (2.10). Expanding (2.9) in terms of
$\Phi$, one finds that
$$\eqalign{
\A_{ij}^{(2)}(\vc\theta,w)&=-{2\over c^2}\int_0^w \d w'\;{f_K(w-w')
f_K(w')\over f_K(w)} \cr
\times & \eck{\Phi_{,ikl}\rund{f_K(w')\vc\theta,w'}\,
x_l^{(1)}(\vc\theta,w')\delta_{kj}
+\Phi_{,ik}\rund{f_K(w')\vc\theta,w'}\A_{kl}^{(1)}(\vc\theta,w') } \;, \cr }
\eqno (A11)
$$
which, after inserting the expressions for $\vc x^{(1)}$ and
$\A^{(1)}$ yields
$$\eqalign{
\A_{ij}^{(2)}(\vc\theta,w)&={4\over c^4}\int_0^w \d w'\;{f_K(w-w')
f_K(w')\over f_K(w)}\int_0^{w'}\d w''\; f_K(w'-w'') \cr
&\times  \biggl[ \Phi_{,ijl}\rund{f_K(w')\vc\theta,w'}\,
\Phi_{,l}\rund{f_K(w'')\vc\theta,w''} \cr
&+{f_K(w'')\over f_K(w')}\Phi_{,ik}\rund{f_K(w')\vc\theta,w'}
\Phi_{,kj}\rund{f_K(w'')\vc\theta,w''}\biggr] \; .\cr }
\eqno (A12)
$$
We consider 1/2 of the trace of $\A^{(2)}$ as the surface mass
density to second order in $\Phi$. Although $\A^{(2)}$ is in general
not symmetric, so that to this order the lens mapping can no longer be
described by an equivalent surface mass density, the asymmetry is
expected to be small in realistic situations, and we shall neglect it
henceforth. Then, after averaging over a source redshift distribution,
as done in (2.15), and replacing the derivatives of $\Phi$ by their
Fourier representation, using the Poisson equation (2.13), one finds
$$\eqalign{
\kappa^{(2)}(\vc\theta)&={9\over 2}\rund{H_0\over c}^4\Omega_{\rm d}^2
\int_0^{w_{\rm H}}\d w'\;g(w')\,{f_K(w')\over a(w')}
\int_0^{w'}\d w''\;{f_K(w'-w'')\over a(w'')} \cr
&\times  \int{\d^3 k'\over (2\pi)^3} {1\over \vert\vec k'\vert^2}
\exp\eck{{{\rm i}\rund{k'_3 w'+f_K(w')\vc
k'\cdot\vc\theta}}} \tilde\delta(\vec k';w') \cr
&\times \int{\d^3 k''\over (2\pi)^3} {1\over \vert\vec k''\vert^2}
\exp\eck{{{\rm i}\rund{k''_3 w''+f_K(w'')\vc
k''\cdot\vc\theta}}} \tilde\delta(\vec k'';w'')\cr
&\times \eck{ {f_K(w'')\over f_K(w')} \rund{\vc k'\cdot\vc k''}^2
+\abs{\vc k'}^2 \vc k'\cdot \vc k''} \; ,  \cr }
\eqno (A13)
$$
where we use the same notation as in the main text, i.e., $\vec k$ has
$\vc k$ as the first two components. 

To calculate the resulting contribution to the skewness according to
(4.1), we note that
$$
\ave{\kappa(\vc\theta_1)\kappa(\vc\theta_2)\kappa(\vc\theta_3)}
\doteq 3
\ave{\kappa^{(1)}(\vc\theta_1)\kappa^{(1)}(\vc\theta_2)
\kappa^{(2)}(\vc\theta_3)}+\O(\Phi^5) \; ,
\eqno (A14)
$$
where the `$\doteq$'-sign means that the two expressions yield the
same result after insertion into (4.1), owing to the symmetry of the
integration there. The correlator in (A14) is calculated by inserting
(A13) for $\kappa^{(2)}$, and (2.15), with $\delta$ replaced by its
Fourier decomposition, for $\kappa^{(1)}$. The resulting expression
then contains a correlation function of four
$\tilde\delta$-terms. Since we consider the lowest, i.e., the linear
order of $\delta$, the four-point correlation function is given by
(4.9). Using arguments very similar to those which led from (4.3) to
(4.5), one finds 
$$\eqalign{
3 & \ave{\kappa^{(1)}(\vc\theta_1)\kappa^{(1)}(\vc\theta_2)
\kappa^{(2)}(\vc\theta_3)}
={243\over 4}\rund{H_0\over c}^8\Omega_{\rm d}^4 
\int_0^{w_{\rm H}}\d w \rund{g(w)\, f_K(w)\,D_+(w)\over a(w)}^2 \cr
&\times \int_0^w\d w'\; {g(w')\,f_K(w')\,f_K(w-w')\,D_+^2(w')\over a^2(w')}
\int{\d^2 k\over (2\pi)^2}\,P_0(k)\,\exp\rund{{\rm i}f_K(w)\vc
k\cdot(\vc\theta_1-\vc\theta_3)} \cr
&\times \int{\d^2 k'\over (2\pi)^2}\,P_0(k')\,\exp\rund{{\rm i}f_K(w')\vc
k'\cdot(\vc\theta_2-\vc\theta_3)} 
\eck{ {f_K(w')\over f_K(w)} {\rund{\vc k\cdot\vc k'}^2\over \abs{\vc
k}^2\abs{\vc k'}^2} +{\vc k\cdot\vc k'\over \abs{\vc k'}^2}}\; .  \cr }
\eqno (A15)
$$
As a final step, we change variables to $\vc s=f_K(w)\vc k$, $\vc
s'=f_K(w')\vc k'$, and insert the resulting expression into (4.1),
using (A14). After applying the definition of $I_\ell$, the
contribution to the skewness from lens-lens coupling and dropping the
Born approximation becomes
$$\eqalign{
\delta\ave{\m^3(\theta)}_{\rm B+C}
&={243\over 8\pi}\rund{H_0\over c}^8\Omega_{\rm d}^4 
\int_0^{w_{\rm H}}\d w \rund{g(w)\,D_+(w)\over a(w)}^2 \cr
&\times  \int_0^w\d w'\;{g(w')\,f_K(w-w')\,D_+^2(w')\over a^2(w')\,f_K(w)} \cr
&\times \int \d^2 s\;P_0\rund{s\over f_K(w)}
\int \d^2 s'\;P_0\rund{s'\over f_K(w')}\cr
&\times \eck{ {\rund{\vc s\cdot\vc s'}^2\over \abs{\vc s}^2\abs{\vc s'}^2}
+{\vc s\cdot\vc s' \over \abs{\vc s'}^2}}
I_\ell(\theta s)\,I_\ell(\theta s')\,I_\ell(\theta\abs{\vc s+\vc
s'})\; . \cr }
\eqno (A16)
$$
We have plotted this correction term, together with
$\delta\ave{\m^3}_g$, in Fig.\ts 9. There we can see that this
correction term is of the same order as that considered in the
previous subsection, i.e., smaller than $\sim 5\%$ on scales larger
than 1\ts arcmin. These corrections are therefore smaller than the
uncertainties introduced by calculating the skewness with quasi-linear
theory. A numerical ray-trace calculation would of course take all the
correction effects mentioned here into account.

It turns out that $\Delta S$ is nearly independent of
cosmology; this can be traced back to the fact that $\delta\ave{\m^3}$
and $\ave{\m^2}^2$ can be expressed as bilinear functions of the
projected power spectrum $P_\kappa$. Assume for a moment that locally,
$P_\kappa(s)$ is a power law; then $\Delta S$ would depend only on
the local slope of this power law, independent of cosmological
factors. Hence, if the local slopes of the power spectra in different
cosmological models are similar, one expects $\Delta S$ to be nearly
independent of the cosmological model.

%

\def\ref#1{\vskip1pt\noindent\hangindent=40pt\hangafter=1 {#1}\par}
\sec{References}
\ref{Bartelmann, M. \& Schneider, P.\ 1991, A\&A 248, 349}
\ref{Bartelmann, M. \& Schneider, P.\ 1994, A\&A 284, 1.}
\ref{Baugh, C., Gazta\~naga, E. \& Efstathiou, G. \ 1995, MNRAS 274, 1049}
\ref{Ben\'\i tez, N. \& Mart\'\i nez-Gonz\'alez, E.\ 1997,
ApJ 477, 27}
\ref{Bernardeau, F., van Waerbeke, L. \& Mellier, Y.\ 1997, A\&A 322,
1 (BvWM)}
\ref{Blandford, R.D. \& Jaroszy\'nski, M.\ 1981, ApJ 246, 1}
\ref{Blandford, R.D., Saust, A.B., Brainerd, T.G. \& Villumsen, J.V.\
1991, MNRAS 251, 600}
\ref{Bonnet, H. \& Mellier, Y.\ 1995, A\&A 303, 331.}
\ref{Bouchet, F., Juszkiewicz, R., Colombi, S. \& Pellat, R.\ 1992, ApJ 394, L5.}
\ref{Bower, R. \& Smail, I.\ 1997, astro-ph/9612151} 
\ref{Colombi, S., Bouchet F.R. \& Hernquist L.\ 1996, ApJ, 465, 14}
\ref{Efstathiou, G.\ 1996, in: {\it Cosmology and large scale
structure}, Les Houches Session LX, R. Schaeffer, J. Silk, M. Spiro \&
J. Zinn-Justin (eds.), North-Holland, p. 133.}
\ref{Fahlman, G., Kaiser, N., Squires, G. \& Woods, D.\ 1994, ApJ 437, 56.}
\ref{Fort, B., Mellier, Y., Dantel-Fort, M., Bonnet, H. \& Kneib,
J.-P.\ 1996, A\&A 310, 705 }
\ref{Fry, J.N. \ 1984, ApJ 279, 499.}
\ref{Gazta\~naga, E. \& Bernardeau, F.\ 1997, astro-ph/9707095}
\ref{Goroff, M.H., Grinstein, B., Rey, S.J. \& Wise, M.B.\ 1986, MNRAS 236, 385.}
\ref{Gunn, J.E.\ 1967, ApJ 147, 61}
\ref{Hamilton, A.J.S., Kumar, P., Lu, E. \& Matthews, A.\ 1991, ApJ
374, L1}
\ref{Jain, B., Mo, H. \& White, S.D.M.\ 1995, MNRAS 276, L25}
\ref{Jain, B. \& Seljak, U.\ 1997, ApJ, in press (JS)}
\ref{Jaroszy\'nski, M.\ 1991, MNRAS 249, 430} 
\ref{Jaroszy\'nski, M.\ 1992, MNRAS 255, 655} 
\ref{Jaroszy\'nski, M., Park, C., Paczy\'nski, B. \& Gott, J.R.\ 1990,
ApJ 365, 22}
\ref{Kaiser, N.\ 1992, ApJ 388, 272}
\ref{Kaiser, N.\ 1995, ApJ 439, L1}
\ref{Kaiser, N.\ 1996, astro-ph/9610120 (K96)}
\ref{Kaiser, N., Squires, G. \& Broadhurst, T.\ 1995, ApJ 449, 460.}
\ref{Kaiser, N., Squires, G., Fahlman, G. \& Woods, D.\ 1994, in: {\it
Clusters of Galaxies}, eds. F.\ts Durret, A.\ts Mazure \& J.\ts Tran
Thanh Van, Editions Frontieres.}
\ref{Lee, M.H.\ \& Paczy\'nski, B.\ 1990, ApJ 357, 32} 
\ref{Luppino, G. \& Kaiser, N.\ 1997, ApJ 475, 20.}
\ref{Mould, J.\ et al.\ 1994, MNRAS 271, 31.}
\ref{Peacock, J.A. \& Dodds, S.J.\ 1996, MNRAS 280, L19}
\ref{Schneider, P.\ 1996, MNRAS 283, 837}
\ref{Schneider, P., van Waerbeke, L., Mellier, Y., Jain, B., Seitz,
S. \& Fort, B.\ 1997, astro-ph/9705122}
\ref{Seitz, S., Schneider, P. \& Ehlers, J.\ 1994,
Class. Quant. Grav. 11, 2345}
\ref{Smail, I., Hogg, D.W., Yan, L. \& Cohen, J.G. 1995, ApJ, 449, L105}
\ref{van Waerbeke, L., Mellier, Y., Schneider, P., Fort, B. \& Mathez, G.
\ 1997, A\& A, 317, 303.}
\ref{Villumsen, J.\ 1995, astro-ph/9507007}
\ref{Villumsen, J.\ 1996, MNRAS 281, 369.}
\ref{Wambsganss, J., Cen, R., Ostriker, J.P. \& Turner, E.L.\ 1995,
Science 268, 274}
\ref{Wambsganss, J., Cen, R., Xu, G. \& Ostriker, J.P.\ 1997, ApJ 475,
L81}

\end